\documentclass[aip, cha, reprint]{revtex4-2}
\usepackage{graphicx,multirow,natbib,hyperref}
\usepackage{amsmath,amssymb}
\usepackage{epstopdf}
\hypersetup{colorlinks,citecolor=black,filecolor=black,linkcolor=black,urlcolor=black}

\usepackage{empheq}
\usepackage[parfill]{parskip}
\usepackage[active]{srcltx}

\usepackage{color}
\usepackage{array}
\usepackage{booktabs}
\usepackage{amsfonts}
\usepackage{dsfont}
\usepackage{graphicx}

\newcommand{\beginsupplement}{%
  \setcounter{section}{0}%
  \setcounter{equation}{0}%
  \setcounter{figure}{0}%
  \setcounter{table}{0}%
 \renewcommand{\thesection}{\Roman{section}}%
 \renewcommand{\theequation}{S\arabic{equation}}%
 \renewcommand{\thefigure}{S\arabic{figure}}%
 \renewcommand{\thetable}{S\arabic{table}}%
}

\begin{document}

\title{Coherence enhanced by {\color{black}{detrained}} oscillators: Breaking $\pi$-reflection symmetry}

\author{Hyunsuk Hong}

\affiliation{Department of Physics and Research Institute of Physics and Chemistry, Jeonbuk National University, Jeonju 54896, Korea}

\author{Jae Sung Lee}
\affiliation{School of Physics, Korea Institute for Advanced Study, Seoul 02455, Korea}

\author{Hyunggyu Park}
\email{hgpark@kias.re.kr}
\affiliation{Quantum Universe Center, Korea Institute for Advanced Study, Seoul 02455, Korea}

\date{\today}

\begin{abstract}
We study a generalized Kuramoto model in which each oscillator carries two coupled phase variables, representing a minimal swarmalator system. Assuming perfect correlation between the intrinsic frequencies associated with each phase variable, we identify a novel dynamic mode characterized by bounded oscillatory motion that breaks the $\pi$-reflection symmetry. This symmetry breaking enhances global coherence  and gives rise to a non-trivial mixed state, marked by distinct degrees of ordering in each variable. Numerical simulations confirm our analytic predictions for the full phase diagram, including the nature of transition. Our results reveal a fundamental mechanism through which {\color{black}{detrained (dynamic)}} oscillators can promote global 
synchronization, offering broad insights into coupled dynamical systems beyond the classical Kuramoto paradigm.
\end{abstract}
\maketitle

\begin{quotation}

Collective synchronization is a hallmark of many complex systems, ranging from flashing fireflies and applauding audiences to neuronal networks and power grids. The Kuramoto model has long served as a fundamental theoretical framework for understanding how simple oscillators achieve synchronization through mutual interactions. In this work, we extend this classical paradigm by considering oscillators that possess two coupled phase variables, providing a minimal mathematical description of so-called swarmalators-entities that both move and synchronize. {\color{black}{ We uncover a previously unreported dynamical regime in which oscillators exhibit bounded oscillations, leading to spontaneous breaking of the $\pi$-reflection symmetry of detrained oscillators. This broken symmetry enhances global coherence and gives rise to an interesting mixed state where the two phase variables display distinct degrees of order. This mechanism provides new insight into how coupled internal and external degrees of freedom shape collective dynamics, with potential relevance to biological swarms, active matter, and
coupled electronic oscillators.
}}

\end{quotation}


\section{Introduction}

The study of coupled oscillators has long been central to understanding collective behavior in complex dynamical systems~\cite{Winfree,Crawford,Kuramoto,Strogatz,StrogatzBook,Pikovsky,Daido,Acebron}, with
real-world applications ranging from biological rhythms~\cite{Buck,Peskin,Gray,StrogatzCircadianClock,Hopfield,Kirst} to synchronization in social networks~\cite{Ott,Arenas,Neda}.
The Kuramoto model~\cite{Kuramoto} and its variants have provided deep insights into how
global coherence can emerge against intrinsic disorder~\cite{KM_Kij,KM_bimodal,KM_Ki,KM_local,KM_thermal}.
Recently, more attention has been paid to systems that exhibit both synchronization and spatial self-organization, such as ``swarmalators''~\cite{KHS17}.
A minimal extension of the Kuramoto model for such systems
considers oscillators with  two coupled variables; an internal phase and a secondary (often spatial) variable, both evolving on a periodic domain.  Interactions between these variables lead to
mutual reinforcement between spatial aggregation and phase synchronization, giving rise to a variety of long-term collective states. These include synchronized clusters, phase waves, and mixed states characterized by strong ordering in one variable and weak ordering in the other~\cite{Kevin22,Yoon22,Hong23}.
 However, the mechanisms underlying such states, particularly the emergence of mixed states, remain poorly understood. In this work,  we explore a simplified yet physically insightful limit of the swarmalator model, in which the intrinsic frequencies associated with the two phase variables 
are perfectly correlated. This assumption retains essential features of the coupling while allowing an analytically tractable framework.

Our most notable finding is the identification of a new dynamical mode; oscillators that are not phase-locked to a fixed point but instead exhibit bounded oscillations with zero mean velocity. Crucially,
these oscillators contribute to collective orderings through interphase coupling.
Using a perturbative approach, we quantify their impact on the order parameters and
show that their dynamics break the $\pi$-reflection symmetry in the phase distribution. This symmetry breaking results in a nonzero contribution from {\color{black}{detrained (dynamic)}} oscillators
to the order parameter--a feature absent in the standard Kuramoto model~\cite{Kuramoto}. Our findings provide deeper insight into general coupled oscillator systems including swarmalator systems, by highlighting the critical roles of coupling asymmetry, frequency correlations, and dynamic entrainment in shaping collective behavior.


\section{Model}
We consider a system of $N$ Kuramoto oscillators with two phase variables. The dynamics of these variables are governed by the coupled differential equations:
\begin{align}
    \dot{x_i} &= v_i+\frac{J}{N} \sum_{j=1}^N \sin(x_j - x_i) \cos(\theta_j - \theta_i), \label{eq:x} \\
    \dot{\theta_i} &= \omega_i + \frac{K}{N} \sum_{j=1}^N \sin(\theta_j - \theta_i ) \cos(x_j - x_i ), \label{eq:theta}
\end{align}
where $x_i$ and $\theta_i$ are the phase variables of oscillator $i$
 $(i=1,\ldots,N)$, each with period $2\pi$, accompanied by intrinsic frequencies $v_i$ and $\omega_i$ drawn randomly from given distributions.
Each variable evolves according to the Kuramoto-type dynamics, but with a coupling strength modulated by the difference in the other
variable. This cross-modulated interaction promotes a mutual enhancement of
local synchrony of both phases for sufficiently large and positive $J$ and $K$.

By interpreting $x_i$ as the position of the $i$-th oscillator, Eqs.~\eqref{eq:x} and \eqref{eq:theta}
effectively describe the dynamics of mobile oscillators with the internal phase $\theta_i$ on a one-dimensional (1D) ring
with length $2\pi$. Phase synchrony in the $x_i$ variables corresponds to spatial aggregation
(swarming) of oscillators, which is enhanced by synchrony in the internal phases.
Conversely, the tendency toward phase synchronization is amplified by spatial clustering of
oscillators.
These mechanisms capture the hallmark behavior of swarmalator systems, i.e.~the co-emergence of
spatial structure and phase coherence.

This model, initially studied in~\cite{Kevin22,Yoon22} and later with
thermal noise replacing quenched intrinsic frequencies~\cite{Hong23},
exhibits several long-term states, including incoherent, phase wave, synchronized, and mixed states.
In this paper, we go beyond the identification of these states to investigate the
underlying mechanisms that govern the behavior of both {\color{black}{entrained (static) and detrained (dynamic)}} oscillators, as well as
their distinct roles in shaping the collective dynamics of the system.

{\color{black}{The model studied in this work is primarily mathematical in nature. Nevertheless, it can serve as a minimal representation of certain real systems, such as population of calling frogs or Janus particles navigating pseudo-one-dimensional grooves or channels. Although this study presents a toy model for such systems, it offers a useful framework for reproducing and elucidating various intriguing phenomena observed in real-world settings~\cite{Kevin22}.
}}


\section{Correlated intrinsic frequencies}
In previous studies, the intrinsic frequencies $v_i$ and $\omega_i$
were drawn independently, implying no correlation between them.
Although some degree of correlation may exist in real-world systems, such correlations are generally expected to have limited influence on the system's collective behavior, except for transition thresholds and potentially critical scalings. Here, we explore the extreme case of perfect correlation by setting $v_i=\omega_i$ for all $i$.
This simplifying assumption not only allows for more tractable analytical treatments but also provides
insight into the behavior of the original, more general model.
For mathematical convenience, we assume the intrinsic frequencies follow a symmetric Lorenzian distribution, $g(\omega)=\frac{\gamma}{\pi}\frac{1}{\omega^2 + \gamma^2}$, centered at zero with width $\gamma$.

The global ordering typically measured by the standard Kuramoto order parameter~\cite{Kuramoto} fails to capture the collective behavior in this model~\cite{Kevin22,Yoon22,Hong23}.
Instead, a more suitable order parameter is the correlation between $x_i$ and $\theta_i$, which effectively characterizes the collective states.
This observation naturally motivates a reformulation of the dynamic equations in terms of  new variables that explicitly encode the correlations between position and phase.
In this context, by introducing the variables
$X_i=x_i + \theta_i$ and $Y_i=x_i-\theta_i$, Eqs.~\eqref{eq:x} and \eqref{eq:theta} can be rewritten as
\begin{align}
\dot{X}_i &= 2\omega_i + J_{+} S_{+} \sin(\Phi_{+}-X_i) + J_{-} S_{-} \sin(\Phi_{-}-Y_i),\label{eq:X}\\
\dot{Y}_i &= J_{-} S_{+} \sin(\Phi_{+}-X_i) + J_{+} S_{-} \sin(\Phi_{-}-Y_i), \label{eq:Y}
\end{align}
where $J_{\pm}=\frac{J\pm K}{2}$ and
$S_{\pm}$ are the magnitudes of the complex order parameters
$Z_{\pm}$, defined as
\begin{align}
Z_{\pm} \equiv S_{\pm} e^{i\Phi_{\pm}} = \frac{1}{N}\sum_{j =1}^N e^{i X_j(Y_j)}. \label{eq:Spm}
\end{align}
Here, $S_{\pm}$ $(\geq 0)$ quantify the degree of coherence in the $X$ and $Y$ variables, respectively,
while $\Phi_{\pm}$ denote the corresponding mean phases. By assuming that $S_\pm$ and $\Phi_\pm$
approach time-independent constants in the long-time limit, as expected for the symmetric $g(\omega)$, we may, without loss of generality, set $\Phi_\pm=0$.

As in the original model~\cite{Yoon22}, we expect
various long-term collective states to emerge:
(a) An incoherent (disordered) state with $(S_+, S_-)=(0,0)$, (b) a phase wave state with $(0,S)$
or $(S,0)$, (c) a mixed state with $(S_1, S_2)$, where $S_1\neq S_2$ and both are finite, and
(d) a synchronized (ordered) state with $(S,S)$.
Due to the introduction of the correlation between $v_i$ and $\omega_i$,
the symmetry between the $X$ and $Y$ variables is explicitly 
{\color{black} broken~\cite{corr}}. As a consequence,
the internal symmetry between $S_+$ and $S_-$, which was present in
the original model, is no longer preserved.
In fact, the synchronized state with $(S,S)$ is not realizable except in special limiting cases.
On the other hand, the absence of intrinsic frequency in the $Y$ dynamics (a consequence of the perfect correlation) renders the mathematical analysis of the system significantly more tractable.

When $J_{-}=0$ (i.e., $J=K$), the dynamics of $X$ and $Y$  become  decoupled, as seen in
Eqs.~\eqref{eq:X} and \eqref{eq:Y}. The $X$ variable follows the standard Kuramoto dynamics with
intrinsic frequency $2\omega$ and coupling strength $J_+$. As is well known~\cite{Kuramoto}, the oscillators are divided into two distinct groups in the long-time
limit; {\color{black}{entrained (static)}} oscillators, which settle into fixed points with $\dot{X}_i=0$ and
{\color{black}{detrained (dynamic)}} oscillators, which circulate incessantly with a nonzero mean velocity.
It is noteworthy that only the static oscillators contribute to the
order parameter $S_+$. In contrast, the $Y$ variable, having zero intrinsic frequency, evolves
according to the Watanabe-Strogatz (WS) dynamics~\cite{WS}. In this case, for any $J_+>0$, all oscillators become static and thus contribute fully to the order parameter $S_-$.
The long-time behavior of the order parameters are then given as follows: $(S_+,S_-)=(0,0)$ for $J_+<0$ (incoherent), $(0,1)$ for $0< J_+< 4\gamma$ (phase wave), and $(S,1)$
with $S=\sqrt{1-4\gamma/J_+}$ for $J_+ > 4\gamma$ (mixed).
Thus, the system exhibits a discontinuous jump in $S_-$ from $0$ to $1$ at $J_+=0$, and a continuous transition in $S_+$ at $J_+=4\gamma$.

When $J_{-}\neq 0$ (i.e., $J\neq K$), on the other hand, the
dynamics of $X$ and $Y$ are coupled, leading to a nontrivial interplay between
the orderings in the $X$ and $Y$ variables.
We note that the system retains a symmetry under the
transformation of $J_-\rightarrow -J_-$ and therefore restrict our analysis to  the case of $J_-\geq 0$, from now on, without loss of generality.

\begin{figure}[!htp]
   \centering
   \includegraphics[width=0.85\columnwidth]{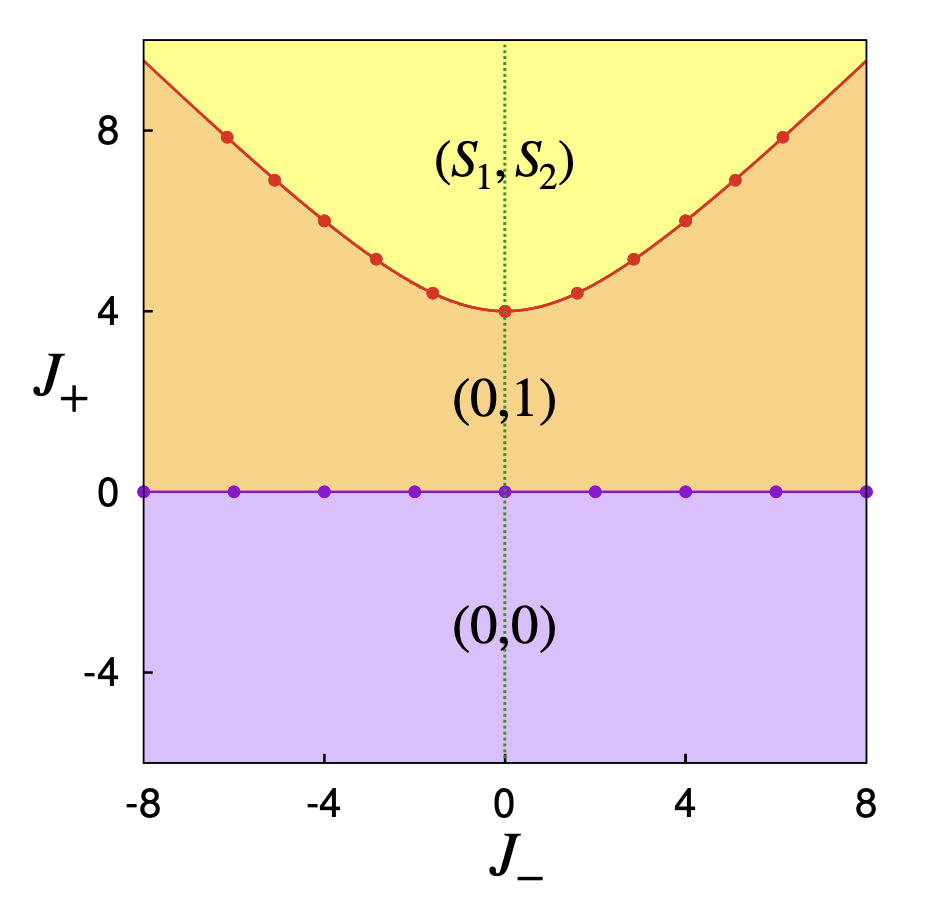}
   \caption{(Color online)
Phase diagram in the $(J_-,J_+)$ plane for $\gamma=1$.
Purple and red circles represent numerical results from simulations, while solid
lines indicate theoretical predictions.
The red curve corresponds to the hyperbolic transition line given by Eq.~(13).
}
  \label{fig:phd}
\end{figure}

Let us first consider the $S_+=0$ solution. In this case, the $Y$ dynamics is unaffected by $X$, even when $J_-\neq 0$, resulting in $S_-=1$ with $Y_i=\Phi_-$ for $J_+>0$, and $S_-=0$ for $J_+<0$.
The $X$ dynamics then simplifies to $\dot{X}_i=2\omega_i$, leading to  $S_+=0$ for all values of $J_+$, thus ensuring the self-consistency of the solution.
As usual, the $(0,1)$ solution for $J_+>0$ is expected to lose stability against solutions with nonzero $S_+$
as $J_+$ increases further beyond a certain threshold. In that regime, the $Y$ dynamics
becomes influenced by the $X$ variable through the emergence of nonzero $S_+$ (see Eq.~\eqref{eq:Y}), which in turn reduces the $Y$ ordering  ($S_-< 1$). This reduction feeds back into the $X$ dynamics, further altering $S_+$, and the cycle continues. This  mutual feedback eventually settles into a
new steady state characterized by $S_+>0$ and $S_- < 1$.
As $J_-$ increases, the coupling between $X$ and $Y$ strengthens, further disrupting
the ordering in $X$. Accordingly, the onset of nonzero $S_+$ is delayed compared to the $J_-=0$ case. This overall qualitative picture is consistent with
the phase diagram shown in Fig.1, which will be analytically derived and numerically confirmed later.

Most interestingly, for finite $S_+$, the $Y$ variable may exhibit a new type of dynamic
behavior, neither static (i.e., fixed point) nor fully dynamic with nonzero mean velocity. For small $S_+$, the solution of Eq.(4) would involve a weak periodic modulation in time, induced by the $X$ dynamics, around the mean phase angle 
$\Phi_-$ (see Eqs.~(S36), (S46), and (S50) of the Supplementary Material (SM)). This time-dependent modulation
persists even in the long-time limit, particularly for oscillators with sufficiently large  $\omega_i$, where $X_i$ continues to evolve with a finite mean velocity.
These oscillators are not static  in the usual
sense with respect to the  $Y$ variable, since they do not converge to a fixed point, but
their  mean phase velocity vanishes, as $Y_i$ does not circulate but instead undergoes bounded oscillations around $\Phi_-$. This type of bounded dynamic motion leads to a reduction in the $Y$ ordering ($S_-<1$). However, the effect on the $X$ ordering is more subtle and nontrivial, as it feeds back through the coupling and alters the collective dynamics in a more intricate manner.


\section{Analytic results}

To investigate the influence of this new dynamic mode on the order parameters,
we employ a perturbation approach, treating $S_+$ as a small parameter near the transition.
As a first step, we decompose the contributions to the order parameters into two components,
arising from static  and dynamic oscillators, respectively.
For convenience, the newly identified dynamical mode, characterized by bounded
oscillations, is included in the dynamic contribution.
This decomposition is expressed as:
\begin{align}
S_{\pm} &= S_{\pm}^{s}+S_{\pm}^{d} \nonumber\\
&= \frac{1}{N} \sum_{j\in \Lambda_s} e^{iX_j(Y_j)} + \frac{1}{N}\sum_{j\in \Lambda_d} e^{iX_j(Y_j)},
\label{eq:Sps_Spd}
\end{align}
where $\Lambda_{s}$ and $\Lambda_{d}$ denote the sets of static and dynamic oscillators, respectively.
Static oscillators, characterized by a stable fixed point ($\dot{X}_i=0$ and $\dot{Y}_i=0$),
are described by
\begin{align}
X_i = \sin^{-1}\bigg(\frac{\omega_i}{aS_+}\bigg),~~\text{and}~~
Y_i = - \sin^{-1}\bigg(\frac{\omega_i}{bS_-}\bigg),
\label{eq:XYs}
\end{align}
with $a=\frac{J_{+}^2-J_{-}^2}{2J_{+}}$ and $b=\frac{J_{+}^2-J_{-}^2}{2J_{-}}$,
and the $\sin^{-1}$ function is restricted to the first quadrant $[0,\pi/2]$.
This fixed-point solution exists only when  both conditions, $|\omega_i/a| \leq S_{+}$ and
$|\omega_i/b| \leq S_-$, are satisfied. Moreover,
a stability analysis requires $a$, $b>0$, i.e., $J_+>J_-$~\cite{ref:SM}.
Since we expect $aS_{+} \leq b S_{-}$ in the small $S_+$ regime (with $S_-\lesssim 1$), the
constraint for the existence of fixed point solutions simplifies to  $|\omega_i| \leq a S_{+}$.
In the limit $N\rightarrow \infty$, the static contribution to the $X$ ordering is
given by~\cite{ref:SM}
\begin{align}
S_{+}^s &= \int_{- aS_{+}}^{aS_{+}} e^{i\sin^{-1}[\omega/(a S_{+})]} g(\omega)d\omega  \nonumber\\
        &= \frac{\gamma}{a S_{+}}\bigg[\sqrt{1+\bigg(\frac{a S_{+}}{\gamma}\bigg)^2}-1\bigg] \nonumber\\
        &= \frac{1}{2}\bigg(\frac{a S_{+}}{\gamma}\bigg)-\frac{1}{8}\bigg(\frac{a S_{+}}{\gamma}\bigg)^3+{\cal O}(S_{+}^5),
\label{eq:Sps}
\end{align}
where we used $g(\omega)=\frac{\gamma}{\pi}\frac{1}{\omega^2 + \gamma^2}$. As expected,
the imaginary part vanishes. A similar expression for $S_-^s$ can be derived
as $S_-^s =\frac{2}{\pi}\big(\frac{aS_+}{\gamma}\big)+{\cal O}(S_+^3)$~(see Eq.~(S93) in SM).

To calculate the contribution from dynamic oscillators, $S_{+}^d$,
we consider the average probability distribution $P_{\omega}^d(X)$ for a given
intrinsic frequency $\omega$ in the long-time limit.
This leads to the following expression:
\begin{align}
S_{+}^d = \int_{|\omega|>a S_{+}} d\omega g(\omega) \int_{-\pi}^{\pi} dX e^{iX}P_{\omega}^d(X).
\label{eq:Zpd}
\end{align}
Deriving the exact form of $P_{\omega}^d(X)$ is generally intractable.
However, in the small-$S_+$ limit, a perturbative expansion becomes possible.
In this regime, $S_-$ remains close to unity, so we set $S_{+}=\varepsilon$
with $\varepsilon<<1$ and approximate
$S_{-}\approx 1-d_1 \varepsilon-d_2 \varepsilon^2$, with constants $d_1$ and $d_2$
to be determined. Expanding all relevant terms up to
${\cal{O}}(\varepsilon^2)$, we derive an analytical expression for $P_{\omega}^d(X)$, which is
rather complicated (see the SM for the explicit expression).

\begin{figure}[!htp]
   \centering
   \includegraphics[width=0.9\columnwidth]{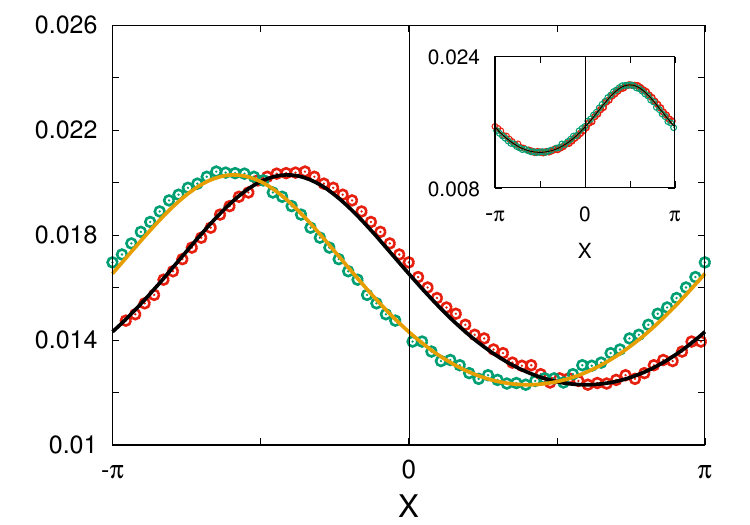}
\caption{(Color online) \label{fig:PDF_X}
{\color{black}{
The $\pi$-reflection symmetry breaking of the average probability distribution function $P_{\omega}^d(X)$ for a detrained (dynamic) oscillator. 
The data (red open circles) are obtained 
from numerical simulations with $N=819,200$, using the parameters $J=9$, $K=3$, and $\omega=4.37$ $(> aS_{+})$, showing excellent agreement with the theoretical prediction (black solid line) given by Eq.~(S59) in  SM~\cite{ref:SM}. For comparison, the same data are replotted with respect to $\pi-X$ (green open circles and orange line), clearly demonstrating the breaking of the $\pi$-reflection symmetry: $P_{\omega}^d(X) \neq P_{\omega}^d(\pi-X)$.
The inset shows the corresponding results for $J=K=5$ ($J_-=0$), where
the $\pi$-reflection symmetry is preserved.
    \label{fig:PDFX}
}}
}
\end{figure}

A key feature of $P_{\omega}^d(X)$, arising from the coupling with the $Y$ variable,
is the breaking of the $\pi$-reflection symmetry:
\begin{align}
P_{\omega}^d(X)\neq P_{\omega}^d(\pi-X)~,
\label{eq:breaking}
\end{align}
indicating that $P_{\omega}^d(X)$ is not purely a function of $\sin(X)$, but also contains $\cos(X)$-like components (see
Eqs.~(S59) and (S52) in SM).
These cosine-like terms originate from
the influence of the new dynamic mode of $Y$ variables,
mediated  by the coupling term when $J_-\neq 0$.
In contrast, this symmetry is preserved in the standard Kuramoto model
resulting in a vanishing real part of the dynamic contribution to
the order parameter in Eq.~\eqref{eq:Zpd}; $\int_{-\pi}^{\pi} dX \cos (X) P_{\omega}^d(X)=0$.
In our case, however, the symmetry breaking leads to a nonzero real contribution to the order parameter. This symmetry breaking is primarily exhibited by slightly detrained oscillators (see SM).
{\color{black}{
The numerical evidence of the $\pi$-reflection symmetry breaking 
(Eq.~\eqref{eq:breaking}) is presented in Fig.~\ref{fig:PDF_X} for a typical
detrained (dynamic) oscillator with $\omega>aS_+$.
}}

After a lengthy calculation~\cite{ref:SM},
we obtain
\begin{align}
S_{+}^d \approx  c_1 S_{+} -c_2 S_{+}^2,
\label{eq:Spd}
\end{align}
with non-negative coefficients given by $c_1=\frac{J_{-}^2}{2J_{+}(J_{+}+2\gamma)}$ and
$c_2=\frac{4a{J_{-}^2}}{3\pi\gamma{J_{+}^2}}$. This result clearly demonstrates that {\em the ordering is
enhanced by dynamic oscillators}, at least within the small-$S_+$ regime.
Combining the contributions from both static and dynamic oscillators, we arrive at
the following self-consistent equation as
\begin{align}
S_{+} = \bigg(\frac{a}{2\gamma}+c_1\bigg) S_{+}-c_2 S_{+}^2 +
{\cal{O}}(S_+^3). \label{eq:Sptotal}
\end{align}
This equation always admits the trivial solution $S_+=0$ and a nontrivial solution emerges
when the linear coefficient satisfies $\frac{a}{2\gamma}+c_1-1>0$.
The nontrivial solution is always stable against the trivial one, so the transition from the $(0,1)$ (phase wave) to the $(S_1,S_2)$ (mixed) state occurs at the critical line given by $\frac{a}{2\gamma}+c_1-1=0$, which yields the following hyperbolic transition line in the $(J_-,J_+)$ plane as
\begin{align}
(J_+-\gamma)^2-J_-^2=(3\gamma)^2~. \label{eq:Jpc}
\end{align}
For comparison, in the absence of dynamic contributions (i.e., setting $c_1=0$),
the corresponding transition line is given by
$(J_+-2\gamma)^2-J_-^2=(2\gamma)^2$. Since the actual transition line always lies
below this reference curve, we conclude that the presence of dynamic oscillators  enhances
the ordering in $X$, resulting in an {\em earlier onset of synchronization}.
It is noteworthy that the self-consistent equation includes a quadratic term $S_+^2$, which
is absent in the standard Kuramoto model. 
As a result, the order parameter $S_+$ grows
linearly near the transition point, characterized by the order parameter exponent $\beta_+=1$, in contrast to the Kuramoto model 
where $\beta=1/2$.

The order parameter $S_-$, associated with the $Y$ variable, can be evaluated in a  similar
manner~\cite{ref:SM} and is given by
\begin{align}
S_{-}=S_-^{s} + S_-^d =1-\frac{c_1}{2} S_+^2 +{\cal O}(S_+^3)~,
\end{align}
where the linear terms from static and dynamic contributions cancel out exactly.
As $S_-$ begins to deviate from 1 (perfect ordering) due to the onset of a nonzero $S_+$, the corresponding transition occurs along the same transition line given by Eq.~\eqref{eq:Jpc}. The reduction in $S_-$ is proportional to $S_+^2$, corresponding to $d_1=0$ and $d_2=\frac{c_1}{2}$, which implies a quadratic decay near the transition with the exponent $\beta_-=2$.

{\color{black}{
We note that the nature of the transition depends on the 
characteristics of the frequency distribution $g(\omega)$, as 
in the conventional Kuramoto model. 
As long as the distribution is symmetric and 
unimodal, such as the Gaussian or Lorentzian forms 
considered in this study, the transition nature is expected to be universal. 
}}

\section{Numerical simulations}
We perform numerical simulations to support the analytic results.
The self-consistent equations, Eqs.~\eqref{eq:X} and \eqref{eq:Y}, are integrated
iteratively, using the order parameters defined by Eq.~\eqref{eq:Spm}.
The system is initialized with random phase values, and time integration is
carried out using the Heun's method~\cite{Heun} for $M_t=10^5$ time steps with a
discrete interval $dt=0.01$. To ensure the system reaches a steady state, the first half
of the simulation ($M_t/2$) is discarded, and the order parameters $S_{\pm}$ are computed by averaging over the remaining time steps. The total number of oscillators is set to $N=10^5$, and we use $\gamma=1$.

\begin{figure}[!htp]
   \centering
    \includegraphics[width=0.9\columnwidth]{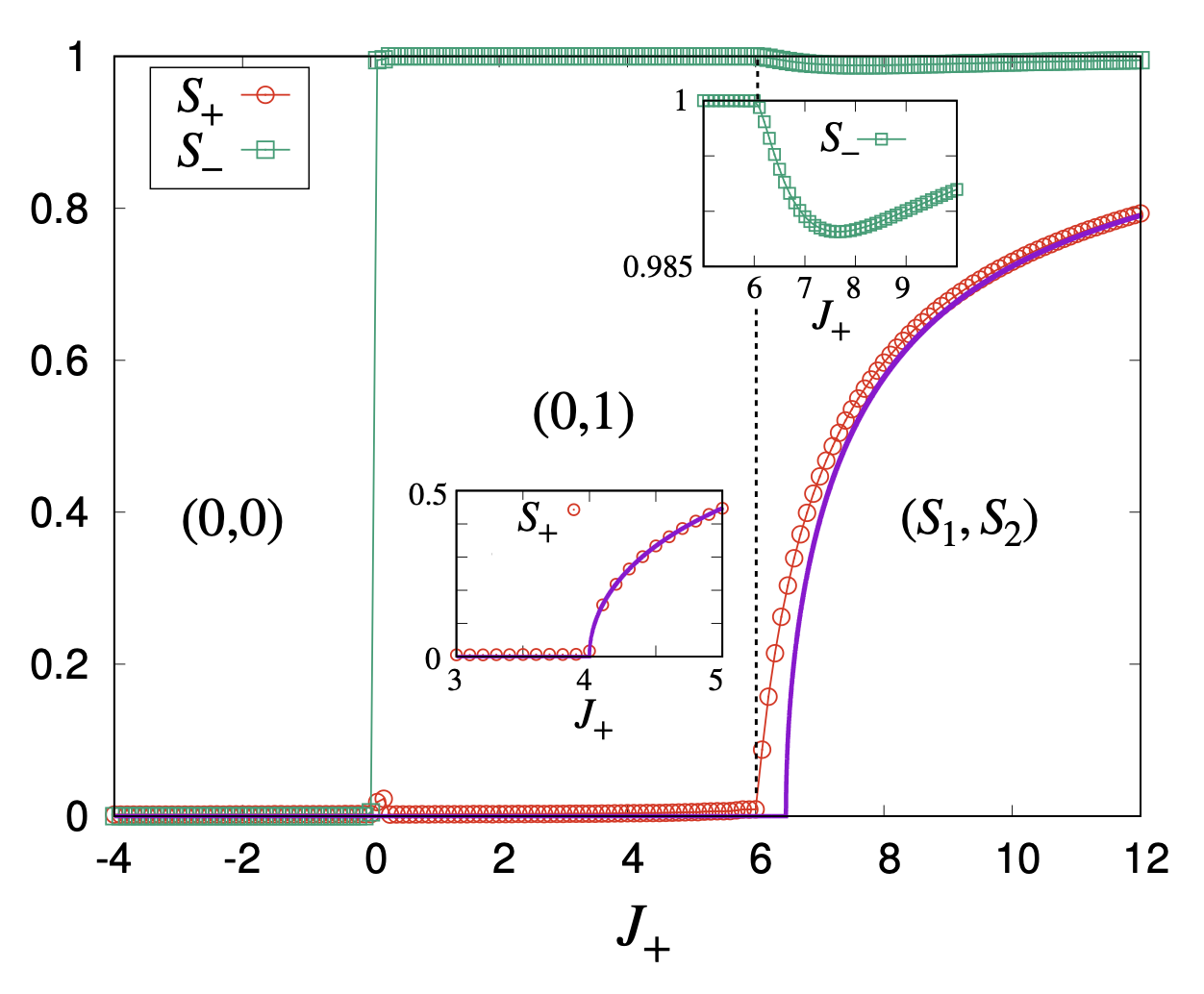}
    \caption{(Color online) The order parameters $S_{\pm}$ are plotted as functions of $J_+$
    for $J_-=4$ and $\gamma=1$.
Open red circles and green squares represent numerical data for $S_{+}$ and $S_{-}$, respectively,
in excellent agreement with analytical predictions, including transition points and their nature.
The order parameter $S_-$ exhibits a discontinuous jump from 0 to 1 at $J_+=0$, and remains nearly constant at 1 thereafter, except for a slight dip beginning around  $J_+ = 6$ (see upper inset). At this point, $S_+$ begins to grow continuously. Both $S_{+}$ and $S_{-}$ approach $1$
asymptotically as $J_+\rightarrow\infty$, without crossing.
The solid purple line corresponds to the prediction obtained by
neglecting dynamic contributions, which clearly deviates from the numerical data.
For comparison, the lower inset shows results for the case with $J_-=0$.
Vertical lines indicate the boundaries between different regimes: $(0,0)$, $(0,1)$, and $(S_1,S_2)$.
  }
    \label{fig:Spm}
\end{figure}

Figure~\ref{fig:Spm} shows the behavior of $S_{\pm}$ as a function
of $J_+$ for a fixed value of $J_-=4$.
The analytic results predict an abrupt transition from the $(0,0)$ state to the $(0,1)$ state at $J_+=0$, followed by a continuous transition to the $(S_1, S_2)$ state  at $J_+=\gamma+\sqrt{J_-^2+ (3\gamma)^2}=6$ (see Eq.~\eqref{eq:Jpc}). These predictions are
in excellent agreement with numerical results. Furthermore, the critical behaviors of
the order parameters, characterized by the exponents $\beta_+=1$ and
$\beta_-=2$, are also confirmed numerically (see Fig.~S2 in SM).
Importantly, neglecting the contribution from dynamic oscillators leads to incorrect
threshold and exponent values (see Eq.~(S30) in SM), as indicated by the purple line in Fig.3.
In the special case $J_-=0$, dynamic contributions vanish and
the purple line agrees with numerical data, yielding the expected exponent $\beta_+=1/2$,
as illustrated in the inset of Fig.3.


\section{Conclusion}
We investigated a population of Kuramoto oscillators
with two coupled phase variables, representing a  minimal model for more general swarmalator systems.
By assuming a perfect correlation between the intrinsic frequencies associated with each phase variable,
we obtain a simplified yet analytically tractable version of the model that retains essential physical features.

A central result of our study is the identification of a novel dynamic mode characterized by bounded oscillations, which
induces $\pi$-reflection symmetry breaking in the dynamics of the coupled phase. This
symmetry breaking draws dynamic oscillators into contributing positively to global ordering, thereby
enhancing overall coherence  and lowering the coherence threshold. As a result, the mixed state emerges
prior to the onset of the symmetric synchronized state.

This form of $\pi$-reflection symmetry breaking is reminiscent of
the transition from apolar nematic to polar symmetry
in systems composed of  non-spherical objects, such as liquid crystals~\cite{LC},  active matter~\cite{AM}, and  various biological systems. Our findings suggest that this type of symmetry breaking through inter-variable coupling may serve as a general mechanism
for an earlier onset of ordering in a broad class of coupled dynamical systems. Consistently with this view, preliminary studies
of the original uncorrelated model  also reveal spontaneous $\pi$-reflection symmetry breaking and
the earlier emergence of coherence in the form of mixed states,  preceding the symmetric synchronized state.

\section*{Supplementary Material}

See the supplementary material for additional details supporting this work.

\section*{Acknowledgements}
We thank J. Um for useful discussions at the early stage of this work. This research was supported by the National Research Foundation of Korea (NRF) grant funded by the Korea government (MSIT) (Grants No.~RS-2024-00348768) (H.H.) and No.~2017R1D1A1B06035497 (H.P.), and individual KIAS Grants No. PG064902 (J.S.L.) and QP013602 (H.P.) at the Korea Institute for Advanced Study.


\clearpage
\section*{Supplementary Material}
\addcontentsline{toc}{section}{Supplementary Material}
\beginsupplement






\title{Supplementary Materials: Coherence enhanced by dynamic oscillators: Breaking $\pi$-reflection symmetry}
\author{Hyunsuk Hong}
\affiliation{Department of Physics and Research Institute of Physics and Chemistry, Jeonbuk National University, Jeonju 54896, Korea}

\author{Jae Sung Lee}
\affiliation{School of Physics, Korea Institute for Advanced Study, Seoul 02455, Korea}

\author{Hyunggyu Park}
\email{hgpark@kias.re.kr}
\affiliation{Quantum Universe Center, Korea Institute for Advanced Study, Seoul 02455, Korea}

\date{\today}


\maketitle
\section{Model}
We consider a system of $N$ coupled Kuramoto oscillators with two phase variables, whose
dynamics are governed by
\begin{align}
    \dot{x_i} &= v_i+\frac{J}{N} \sum_{j=1}^N \sin(x_j - x_i) \cos(\theta_j - \theta_i), \label{Seq:x} \\
    \dot{\theta_i} &= \omega_i + \frac{K}{N} \sum_{j=1}^N \sin(\theta_j - \theta_i ) \cos(x_j - x_i ), \label{Seq:theta}
\end{align}
$i=1, \dots, N$, where $x_i$ and $\theta_i$ are the phase variables of oscillator $i$,
each with period $2\pi$.
The parameters $v_i$ and $\omega_i$ denote the intrinsic frequencies associated with
each phase variable, and  are randomly drawn from given distributions.
In this work, we focus on the case of perfect correlation by setting $v_i=\omega_i$  for all $i$.
For convenience, we assume that $\omega_i$ (also $v_i$) follows a
Lorentzian distribution:
\begin{align}
g(\omega)=\frac{\gamma}{\pi}\frac{1}{\omega^2 + \gamma^2}, \label{Seq:gw}
\end{align}
centered at zero with width $\gamma$.

As shown in previous studies (see Refs. [25]-[26] of the main paper), the global ordering
in this system is not adequately captured by the standard Kuramoto order parameters for $x_i$ and $\theta_i$.
This limitation arises from the symmetry of the dynamics, which are invariant
under simultaneous translations
of both $x_i$ and $\theta_i$ by $\pi$  for each $i$. As a result, conventional measures of phase coherence
vanish identically. even when the system exhibits strong correlations.
Instead, the collective states are characterized by the correlation between $x_i$ and $\theta_i$.
To probe this correlation structure, we introduce the following change of variables:
\begin{align}
X_i=x_i + \theta_i,~~{\mbox{and}}~~ Y_i=x_i-\theta_i,
\end{align}
which allows Eqs.~(\ref{Seq:x}) and (\ref{Seq:theta}) to be rewritten as
\begin{align}
\dot{X}_i &= 2\omega_i + J_{+} S_{+} \sin(\Phi_{+}-X_i) + J_{-} S_{-} \sin(\Phi_{-}-Y_i),\label{Seq:X}\\
\dot{Y}_i &= J_{-} S_{+} \sin(\Phi_{+}-X_i) + J_{+} S_{-} \sin(\Phi_{-}-Y_i), \label{Seq:Y}
\end{align}
where $J_{\pm}=\frac{J\pm K}{2}$. The quantities $S_{\pm}$ denote the magnitudes of the complex
order parameters $Z_{\pm}$, defined by
\begin{align}
Z_{+} &\equiv S_{+} e^{i\Phi_{+}} = \frac{1}{N}\sum_{j =1}^N e^{i X_j}, \label{Seq:Sp} \\
Z_{-} &\equiv S_{-} e^{i\Phi_{-}} = \frac{1}{N}\sum_{j =1}^N e^{i Y_j}. \label{Seq:Sm}
\end{align}
Here, $S_{\pm}(\geq 0)$ quantify the degree of coherence in the $X$ and $Y$ variables,
respectively, while $\Phi_{\pm}$ denote the corresponding mean phases.

Assuming that $S_\pm$ and $\Phi_\pm$ converge to time-independent constants in the
long-time limit, as expected for the symmetric distribution $g(\omega)$, we can set $\Phi_{\pm}=0$ without
loss of generality.  This is achieved by shifting the $X$ (or $Y$)
variables by $\Phi_+$ (or $\Phi_-$), effectively eliminating the mean phases.

To explore how dynamic (detrained) oscillators influence the systems's ordering, we first
split all oscillators into two groups: static (entrained) ones and dynamic (detrained) ones.
We call an oscillator $X_i$ (or $Y_i$) {\it{static}}
if $\dot{X}_i = 0$ (or $\dot{Y}_i = 0$) in the long-time limit, and {\it{dynamic}} otherwise.

To quantify each group's contribution separately, we then define the following order parameters:
\begin{align}
S_{\pm} &= S_{\pm}^s+S_{\pm}^d, \label{Seq:Zpm} \nonumber\\
	&=\frac{1}{N}\sum_{j\in\Lambda_{s}}e^{i X_j(Y_j)} + \frac{1}{N}\sum_{j\in\Lambda_{d}}e^{i X_j(Y_j)},
\end{align}
where the superscripts `s' and `d' in the order parameters $S_{\pm}$ denote `static' and `dynamic' oscillator groups, respectively, while $\Lambda_s$ and $\Lambda_d$ represent the
corresponding sets of static and dynamic oscillators.

\section{Case with $J_{-}=0$}
We begin by considering the case $J_{-}=0$ (i.e., $J=K$), which is more analytically
tractable and closely resembles the Kuramoto model, allowing for direct comparisons with previous studies.
In this case, the equations for $X_i$ and $Y_i$ in Eqs.~(\ref{Seq:X}) and (\ref{Seq:Y})
become completely decoupled and take the form
\begin{align}
\dot{X}_i &= 2\omega_i + J_{+} S_{+} \sin(\Phi_{+}-X_i),  \label{Seq:X_sameJK}\\
\dot{Y}_i &= J_{+} S_{-} \sin(\Phi_{-}-Y_i), \label{Seq:Y_sameJK}
\end{align}
$i=1,\cdots,N$.
Note that the equation for $X_i$ reduces to the Kuramoto model with a natural frequency
of $2\omega_i$ and a coupling strength of $J_{+}$.
For simplicity, we omit the index $i$ in what follows, unless it is necessary.
As is well known from previous studies of the Kuramoto model with $J_+>0$, oscillators with frequencies
satisfying $|\omega| \leq \omega_c$, where $\omega_c=J_{+}S_{+}/2$, become
static, i.e., {\it{locked}} (or {\it{entrained}}), while those with $|\omega| > \omega_c$
are dynamic, i.e., {\it{unlocked}} (or {\it{detrained}}). For $J_+< 0$, all oscillators are dynamic.
On the other hand, the dynamics of the $Y$ variable corresponds to the
zero intrinsic-frequency case, known as the Watanabe-Strogatz model (see Ref. [28] of the main paper).

The dynamics of the oscillators in each group can be described as follows:
\vspace{0.1cm}
\begin{itemize}
\item{\bf{Static (entrained) oscillators}}
\end{itemize}
For static oscillators that reach a time-independent state with  $\dot{X}=0$ and $\dot{Y}=0$ in the long-time limit, the steady-state (fixed point) values  are given by
\begin{align}
X &= \Phi_{+} + \sin^{-1}(\omega/\omega_c),~~~{\mbox{for}}~|\omega|\leq {\omega_c} \label{Seq:X_static} \\
Y &= \Phi_{-},~~~
\label{Seq:Y_static}
\end{align}
for $J_+>0$. Note that Eq.~(\ref{Seq:Y_static}) holds for all oscillators, since the  $Y$-dynamic equation  lacks an
intrinsic frequency term. As a result, each oscillator's $Y$ value remains fixed at the mean phase $\Phi_-$, regardless of $\omega$.
For $J_+< 0$, the $X$ dynamics admits no stable fixed points, and hence all oscillators are dynamic.
In contrast, $Y_i$ still approaches a fixed point determined by initial conditions. These fixed points collectively satisfy
a discrete rotational symmetry  such as $\sum_j e^{iY_j}=0$, yielding $S_-=0$ in the long-time limit (see Ref. [28] of the main paper).

\vspace{0.1cm}
\begin{itemize}
\item{\bf{Dynamic (detrained) oscillators}}
\end{itemize}
For dynamic oscillators, satisfying $\dot{X}\neq 0$, the evolution of $X$ follows
\begin{align}
X = X(0)+{\tilde{\omega}} t + f(t),~~~{\mbox{for}}~|\omega| > \omega_c, \label{Seq:X_dynamic}
\end{align}
where $X(0)$ is the initial condition, and $f(t)$ is a bounded  time-dependent function capturing time-dependent fluctuations.
The effective average frequency  ${\tilde{\omega}}$, modified by coupling, is given by
\begin{align}
\tilde{\omega}={\rm sgn} (\omega)\sqrt{(2\omega)^2-(J_{+}S_{+})^2},
\end{align}
where ${\rm sgn}(\omega)$ denotes the sign function, returning $+1$ for $\omega>0$, $-1$ for $\omega<0$, and $0$ for $\omega = 0$.

\subsection{Order parameters for $J_{-}=0$}

From Eq.~(\ref{Seq:Y_static}), it follows  directly that all oscillators share the same fixed point, implying perfect phase
alignment in the $Y$-dynamics for $J_+>0$. Consequently, the corresponding order parameters are given as
\begin{align}
S_{-}^s=1,~~{\mbox{and}}~~S_{-}^d=0~,
\label{Seq:WS}
\end{align}
indicating that all oscillators are static and fully synchronized in $Y$.
In comparison, for $J_+<0$, synchronization is lost entirely, yielding $S_{-}^s=0$, and $S_{-}^d=0$.
Thus, the order parameter $S_-$ exhibits a discontinuous jump from 0 to 1 at $J_+=0$.

We now consider the order parameters $S_{+}^s$ and $S_{+}^d$, associated with the $X$-dynamics for $J_+>0$.
In the $N\rightarrow\infty$ limit, the contribution of static oscillators ($|\omega| \leq \omega_c$) to the order parameter $S_{+}$
is given by
\begin{align}
S_{+}^s = \int_{-\omega_c}^{\omega_c} d\omega g(\omega)
e^{i \sin^{-1}(\omega/\omega_c)}~, \label{Seq:Zps_sameJK_Ninfty}
\end{align}
where $g(\omega)$ is given  in Eq.~\eqref{Seq:gw}.
The imaginary part of the integral in Eq.~(\ref{Seq:Zps_sameJK_Ninfty})
vanishes due to the odd symmetry of the integrand over the symmetric interval. As a result, the expression
reduces to
\begin{align}
S_{+}^s &= \int_{-\omega_c}^{\omega_c} d\omega~g(\omega)~
\sqrt{1-\left(\frac{\omega}{\omega_c}\right)^2} \nonumber\\
&=\sqrt{1+\left(\frac{\gamma}{\omega_c}\right)^2}-\frac{\gamma}{\omega_c},
\label{Seq:Sps_integral_sameJK}
\end{align}
with the distribution width $\gamma$ in  Eq.~\eqref{Seq:gw}.
We note that the expression in the right-hand side of Eq.~(\ref{Seq:Sps_integral_sameJK})
explicitly depends on $S_+$, as $\omega_c=J_{+}S_{+}/2$.

To obtain the total order parameter $S_{+}$, encompassing both static and dynamic oscillators, we must also determine $S_{+}^d$ for the dynamic oscillators.
According to previous studies on the Kuramoto model, it is well established that dynamic oscillators with frequencies $|\omega| > \omega_c$ do not contribute
to phase ordering. For completeness, we briefly explain why these dynamic oscillators fail to contribute, as this will be relevant
to the discussion that follows.

To calculate the contribution from dynamic oscillators, $S_+^d$,  it is convenient to consider  the long-time average probability distribution function  $P_{\omega}^d(X)$
for a given intrinsic frequency $\omega$. In the  $N\rightarrow\infty$ limit,  $S_{+}^d$ can be expressed as
\begin{align}
S_{+}^d = \int_{|\omega|>\omega_c} d\omega g(\omega) \int_{-\pi}^{\pi} dX e^{iX} P_{\omega}^d(X)~,
\label{Seq:Zpd_integral}
\end{align}
where $P_{\omega}^d(X)$  can be derived from the expression for ${\dot{X}}$ given in Eq.~\eqref{Seq:X_sameJK}.
In general, the probability  of finding an oscillator near a given phase $X$ is inversely proportional to its speed $|{\dot{X}}|$;
slower oscillators spend more time within a given interval $dX$, while faster ones pass through more quickly.
Imposing the normalization condition $\int_{-\pi}^{\pi} P_{\omega}^d (X)dX=1$, and using the fact that
the dynamics are $2\pi$-periodic in average for dynamic oscillators, we obtain
\begin{align}
P_{\omega}^d(X)
&=\frac{1}{\dot X} \left( \int_{-\pi}^\pi \frac{dX}{\dot X} \right)^{-1}  \nonumber \\
&=\frac{{\mbox{sgn}}(\omega)}{2\pi}\cdot\frac{\sqrt{(2\omega)^2- (J_{+}S_{+})^2}}{2\omega-J_{+}S_{+}\sin X}. \label{Seq:P_sameJK}
\end{align}
Note that $P_{\omega}^d(X)$ satisfies the $\pi$-reflection symmetry:
\begin{align}
P_{\omega}^d(X)=P_{\omega}^d(\pi-X)~,
\end{align}
reflecting symmetry about $X=\pi/2$.
This immediately implies  $\int_{-\pi}^{\pi} \cos (X) P_{\omega}^d(X)=0$,
causing  the real part of the integral in Eq.~\eqref{Seq:Zpd_integral} to vanish.
The imaginary part also vanishes upon integration over the symmetric $g(\omega)$.
We thus conclude $S_{+}^d = 0$.

Consequently, the order parameter $S_+$ is determined entirely by the static contribution $S_{+}^s$,
i.e.~$S_{+}^s = S_+$.  Thus, Eq.~(\ref{Seq:Sps_integral_sameJK})  becomes a self-consistency equation for $S_+$,
which yields the stable nonzero solution
\begin{align}
S_{+} = S_{+}^s=\sqrt{1-\frac{4\gamma}{J_{+}}}~~~\quad\text{for}~ J_{+} \geq 4\gamma,
\label{Seq:Sps_sameJK}
\end{align}
which defines the threshold value $J_+^c = 4\gamma$. For $J_+<J_+^c$, the only solution is the trivial one, $S_+=0$.
Accordingly, near the transition point, the order parameter $S_{+}$ exhibits the critical behavior
\begin{align}
S_{+} \sim(J_{+}-J_{+}^c)^{\beta_+}, \label{Seq:Sps_beta_sameJK}
\end{align}
with the critical exponent $\beta_+=1/2$, consistent with the standard Kuramoto model for oscillators with natural frequency $2\omega$.

\begin{figure}
   \centering
    \includegraphics[width=0.95\columnwidth]{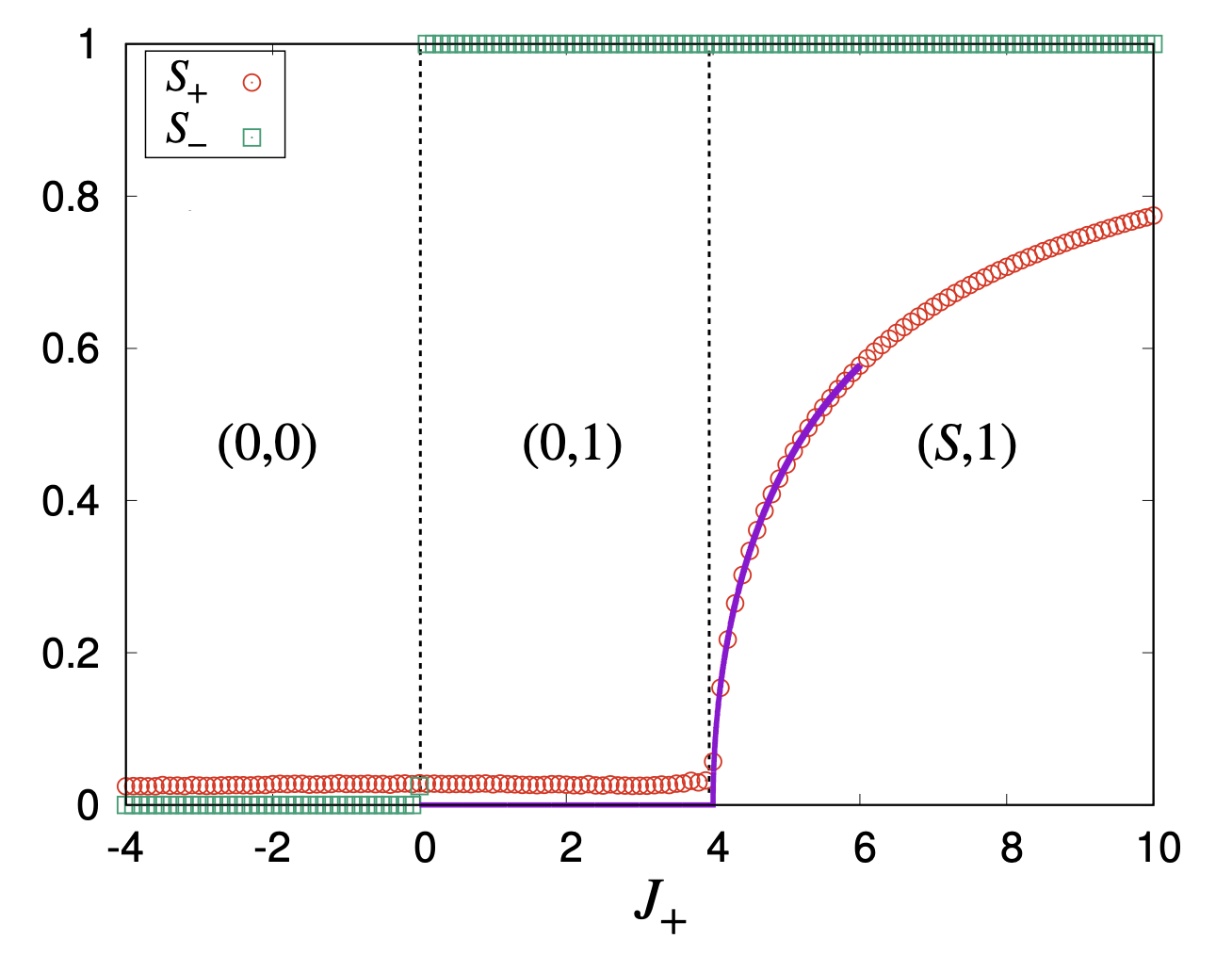}
    \caption{(Color online) The order parameters $S_{\pm}$ are plotted as functions of
$J_{+}$ for fixed $J_{-}=0$ with $\gamma=1$.  Open red circles and green squares represent
numerical data for $S_{+}$ and $S_{-}$, respectively.
Vertical dashed lines at $J_{+}=0$ and $J_{+}=4$ indicate
the boundaries between three distinct states: $(S_{+}, S_{-}) = (0,0)$, $(0,1)$, and $(S,1)$.
The solid purple line denotes the analytic solution for $S_+$ given by
Eq.~(\ref{Seq:Sps_sameJK}).}
    \label{SMfig:Spm_sameJK}
\end{figure}

To confirm this analysis, we perform numerical simulations of the model described by Eqs.~(\ref{Seq:X_sameJK}) and (\ref{Seq:Y_sameJK}) and measured the order parameters $S_{\pm}$.
Figure~\ref{SMfig:Spm_sameJK} shows the behavior of $S_{\pm}$ as a function of $J_{+}$ for fixed
$J_{-}=0$.  Open red circles and green squares represent numerical data for  $S_{+}$ and $S_{-}$, respectively,
both of which show excellent agreement with the analytical predictions: Eq.~\eqref{Seq:Sps_sameJK}
 for $S_+$ (the solid purple line)  and Eq.~\eqref{Seq:WS}.

\section{Case with $J_{-}\neq 0$}
We now investigate the ordering behavior when $J_{-} \neq 0$ (i.e., $J \neq K$). In this regime, the equations for $X_i$ and $Y_i$ in Eqs.~(\ref{Seq:X}) and (\ref{Seq:Y})  become coupled, leading to various nontrivial behaviors.
The presence of nonzero coupling also makes it difficult to obtain general analytic solutions.
Thus, we adopt a perturbative approach, valid in the regime of small $S_+$ and near $S_-\approx 1$,
corresponding to the onset of a nontrivial mixed state $(S_1, S_2)$, as shown in Fig.~3 of the main text.

Before proceeding to a detailed analysis,
we note that the system retains a symmetry under the
transformation $J_-\rightarrow -J_-$,
similar to that of the original uncorrelated model.
This symmetry arises from  the interchange of variables $x\leftrightarrow\theta$
and parameters $J\leftrightarrow K$ in Eqs.~\eqref{Seq:x} and \eqref{Seq:theta}, which
holds even in the presence of correlation between $v_i$ and $\omega_i$.
Equivalently, Eqs.~\eqref{Seq:X} and \eqref{Seq:Y} are invariant under the combined transformation,
$J_-\rightarrow -J_-$ and $Y\rightarrow -Y$ (with $\Phi_-\rightarrow -\Phi_-$), implying
that the order parameter values remain unchanged under the sign change of $J_-$.
We therefore restrict our analysis to  the case of $J_-\geq 0$ without loss of generality.

\subsection{Order parameter $S_{+}$ for $J_{-}\neq 0$}

To evaluate the order parameter $S_+$, it is necessary to calculate $S_+^s$ and $S_+^d$ separately. First, $S_+^s$ can be evaluated by following a similar procedure used for obtaining $S_+^s$ in
the $J_{-}=0$ case. When $J_{-} \neq 0$, the static solution $(X^s, Y^s)$ of Eqs.~\eqref{Seq:X} and \eqref{Seq:Y}, satisfying $\dot{X}=0$ and $\dot{Y}=0$, is expressed as follows:
\begin{align}
X^s &= \sin^{-1}\bigg(\frac{\omega}{aS_+}\bigg)~~~{\mbox{for}}~~|\omega| \leq aS_{+},
\label{Seq:Xs} \\
Y^s &= - \sin^{-1}\bigg(\frac{\omega}{bS_-}\bigg)~~~{\mbox{for}}~~|\omega| \leq bS_{-},
\label{Seq:Ys}
\end{align}
where
\begin{align}\label{Seq:ab}
a=\frac{J_{+}^2-J_{-}^2}{2J_{+}}, \quad\mbox{and}\quad b=\frac{J_{+}^2-J_{-}^2}{2J_{-}}.
\end{align}
The domain of the $\sin^{-1}$ function is restricted to the first quadrant $[0,\pi/2]$, and
a stability analysis requires both  $a>0$ and $b>0$, which corresponds to the condition
$J_+>J_-$~(see the final section of this Supplementary Material).

In the regime of interest (small $S_+$ and large $S_-$), the inequality  $aS_+\leq b S_-$ holds, allowing
the condition for an oscillator with frequency $\omega$ to remain static to be simplified as $|\omega|<aS_+$.
In the limit $N\rightarrow \infty$, the order parameter $S_{+}^{s}$ for the static oscillators
can then be expressed as
\begin{align}
S_{+}^s = \int_{-aS_{+}}^{aS_{+}} e^{i\sin^{-1}[\omega/(aS_{+})]} g(\omega)d\omega. \label{Seq:Zps_JneqK}
\end{align}
Using the Lorentzian distribution~Eq.\eqref{Seq:gw} for $g(\omega)$, the order parameter $S_+^s$ is
given by
\begin{align}
S_{+}^s &= aS_{+}  \int_{-1}^{1} \frac{\gamma}{\pi}\frac{\sqrt{1-x^2}}{(aS_{+}x)^2+\gamma^2} dx \nonumber\\
&= \frac{\gamma}{aS_{+}}\sqrt{1+\bigg(\frac{aS_{+}}{\gamma}\bigg)^2}-\frac{\gamma}{aS_+} \nonumber\\
&= \frac{1}{2}\Bigg(\frac{aS_{+}}{\gamma}\Bigg)-\frac{1}{8}\Bigg(\frac{aS_{+}}{\gamma}\Bigg)^3 + {\cal{O}}(S_+^5),
\label{Seq:Sps_JneqK}
\end{align}
leading to
\begin{align}
	S_+^s = \frac{a S_+}{2\gamma} - \Bigg(\frac{a S_+}{2\gamma}\Bigg)^3 + {\cal{O}}(S_+^5). \label{Seq:Sps_JneqK_approx}
\end{align}

By neglecting the contribution from dynamic oscillators, i.e.~assuming $S_+^d =0$ and setting $S_+ = S_+^s$, we obtain
the nontrivial  solution of Eq.~\eqref{Seq:Sps_JneqK} as
\begin{align}
	S_+ = \sqrt{1-\frac{2\gamma}{a}} ~~~\quad\text{for}~ a \geq 2\gamma~.
	\label{Seq:zeroDynamicS}
\end{align}
Substituting $a=\frac{J_{+}^2-J_{-}^2}{2J_{+}}$, we obtain the transition line for the onset of $S_+$:
\begin{align}
(J_+-2\gamma)^2-J_-^2=(2\gamma)^2.\label{Seq:zero_dyn_tran_line}
\end{align}
In Fig.~3 of the main text, this result (represented by the purple line) clearly deviates from the numerical data,
indicating that the contribution from dynamic oscillators cannot be neglected.

To account for this discrepancy,  we next compute $S_+^d$, which can be evaluated using
the distribution $P_\omega^d (X)$, following a procedure similar to that in the case of $J_{-} = 0$.
Once $P_{\omega}^d(X)$ is determined, $S_{+}^d$ can be obtained from
\begin{align}
S_{+}^d = \int_{|\omega|>aS_{+}} d\omega g(\omega)
\int_{-\pi}^{\pi} dX e^{iX} P_{\omega}^d(X).
\label{Seq:Spd}
\end{align}

We begin  by calculating  $P_{\omega}^d(X)$ using a perturbative expansion valid for small $S_+$ and near $S_-=1$.
All relevant quantities are expanded up to second order in $S_+$ as
\begin{align}
&S_{+}\equiv \varepsilon, \\
&S_{-} = 1-\sigma^{(1)}\varepsilon-\sigma^{(2)}\varepsilon^2 + \cdots, \label{Seq:Sm_epsilon}\\
&X_{\omega} \equiv  X-\Phi_{+}=X_{\omega}^{(0)}+X_{\omega}^{(1)}\varepsilon+X_{\omega}^{(2)}\varepsilon^2 + \cdots, \label{Seq:X_epsilon}\\
&Y_{\omega} \equiv Y-\Phi_{-}=Y_{\omega}^{(1)}\varepsilon +Y_{\omega}^{(2)}\varepsilon^2+\cdots. \label{Seq:Y_epsilon}
\end{align}
Here, $X_\omega$  and $Y_\omega$ denote the phase variables relative to their respective mean phases for an oscillator with intrinsic frequency $\omega$. The coefficients $\sigma^{(1)}$ and $\sigma^{(2)}$ characterize
the linear and quadratic corrections in $S_-$, respectively. As shown later in Eq.~\eqref{Seq:Sm_withSp},
the $O(\varepsilon)$ term in $S_{-}$ vanishes, allowing us to set $\sigma^{(1)} = 0$ from the outset for simplicity and
denote $\sigma^{(2)}\equiv\sigma$.

Substituting Eqs.~\eqref{Seq:Sm_epsilon},~\eqref{Seq:X_epsilon}, and~\eqref{Seq:Y_epsilon} into Eq.~\eqref{Seq:X} and
retaining terms up to $O(\varepsilon^2)$, we obtain
\begin{align}
{\dot{X}}_{\omega} = 2\omega - J_{+}\varepsilon \sin X_{\omega} - J_{-}\sin Y_{\omega} + O(\varepsilon^3),  \label{Seq:Xomega_epsilon}
\end{align}
where $\sin X_{\omega} \approx \sin X_\omega^{(0)} + X_\omega^{(1)} \cos X_\omega^{(0)} \varepsilon$ and $\sin Y_\omega \approx Y_\omega^{(1)} \varepsilon + Y_\omega^{(2)} \varepsilon^2 $.
By collecting terms of the same order in $\varepsilon$ on both sides of Eq.~(\ref{Seq:Xomega_epsilon}), we obtain the 0th-order, 1st-order, and 2nd-order equations as follows:
\begin{align}
&{\dot{X}}_{\omega}^{(0)}=2\omega, ~~(0{\rm th}) ~~\label{Seq:0th} \\
&{\dot{X}}_{\omega}^{(1)}
=-J_{+}\sin X_{\omega}^{(0)}-J_{-}Y_{\omega}^{(1)}, ~~(1{\rm st}) \label{Seq:X_1st} \\
&{\dot{X}}_{\omega}^{(2)}  =-J_{+}X_{\omega}^{(1)}\cos X_{\omega}^{(0)}-J_{-}Y_{\omega}^{(2)}. ~~(2{\rm nd}) \label{Seq:X_2nd}
\end{align}
Similarly, by substituting the Eqs.~(\ref{Seq:Sm_epsilon})-(\ref{Seq:Y_epsilon}) into  Eq.~(\ref{Seq:Y}), we obtain
\begin{align}
	{\dot{Y}}_\omega &= -J_{-}\varepsilon\sin X_{\omega} -J_{+}\sin Y_{\omega} +O(\varepsilon^3). \label{Seq:y1y2}
\end{align}
Collecting terms of the same order in $\varepsilon$ on both  sides of Eq.~(\ref{Seq:y1y2}) leads to
\begin{align}
	&{\dot{Y}}_\omega^{(1)} = -J_{-}\sin X_\omega^{(0)} - J_{+} Y_{\omega}^{(1)},~~(1{\rm st}) \label{Seq:y1}  \\
	&{\dot{Y}}_\omega^{(2)} = -J_{-} X_{\omega}^{(1)} \cos X_\omega^{(0)}- J_{+} Y_{\omega}^{(2)}. ~~(2{\rm nd})
	\label{Seq:y2}
\end{align}

The solution for $X_{\omega}^{(0)}$ can be readily obtained from Eq.~\eqref{Seq:0th} as follows:
\begin{align}
	X_{\omega}^{(0)}=2\omega t \label{Seq:slnX0}
\end{align}
with the initial condition $X_{\omega}(0)=0$ for convenience.
By using Eq.~\eqref{Seq:slnX0}, Eq.~\eqref{Seq:y1} can be converted into the following form:
\begin{align}
{\dot{Y}}_\omega^{(1)}+J_{+}Y_{\omega}^{(1)}=f_1(t), \label{Seq:Y1_diff_eq}
\end{align}
where $f_1 (t)=-J_{-}\sin 2\omega t$. The solution of the first order differential equation~\eqref{Seq:Y1_diff_eq} is
\begin{align}
Y_{\omega}^{(1)} &=e^{-J_{+}t}\int_{0}^t dt^{\prime} e^{J_{+}t^{\prime}}f_1 (t^{\prime}) =\frac{J_{-} \cos(2\omega t + \beta_{\omega}) }{\sqrt{J_{+}^2+4\omega^2}},  \label{Seq:y1solution}
\end{align}
where $\beta_{\omega}$ satisfies the following relations:
\begin{align}
\cos\beta_{\omega} = \frac{2\omega}{\sqrt{J_{+}^2+4\omega^2}} ~~~~{\rm and}~~~~
\sin\beta_{\omega} = \frac{J_{+}}{\sqrt{J_{+}^2+4\omega^2}}. \label{Seq:sincosbeta}
\end{align}
{\color{black}  This solution clearly indicates that the variable $Y_\omega$ 
in Eq.~\eqref{Seq:Y_epsilon} exhibits bounded sinusoidal oscillations around the mean phase, with a small amplitude of order $\varepsilon$, for detrained oscillators.}

Substituting Eq.~(\ref{Seq:y1solution}) into Eq.~(\ref{Seq:X_1st}) yields
\begin{align}
{\dot{X}}_{\omega}^{(1)} = -J_{+}\sin 2\omega t - \frac{J_{-}^2 \cos(2\omega t + \beta_{\omega})}{\sqrt{J_{+}^2+4\omega^2}},
\end{align}
which results in
\begin{align}
X_{\omega}^{(1)}=\frac{J_{+}}{2\omega}\cos 2\omega t - \frac{J_{-}^2 \sin(2\omega t + \beta_{\omega})}{2\omega\sqrt{J_{+}^2+4\omega^2}}   + C_X^{(1)}, \label{Seq:X1solution}
\end{align}
where $C_X^{(1)}$ denotes an integration constant and is set to zero for convenience since its arbitrary value does not affect the final result.

Then, by substituting Eq.~\eqref{Seq:X1solution} into Eq.~(\ref{Seq:y2}) and following the same calculation steps from Eq.~\eqref{Seq:Y1_diff_eq} to \eqref{Seq:sincosbeta}, $Y_\omega^{(2)}$ can be evaluated as
\begin{align}
Y_{\omega}^{(2)}
&=e^{-J_{+}t}\int_0^t dt^{\prime}\left(\frac{-J_{+}J_{-}}{2\omega}\right)e^{J_{+}t^{\prime}}\cos^2 2\omega t^{\prime}
\nonumber\\
&+ e^{-J_{+}t}\int_0^t dt^{\prime} \frac{J_{-}^3 e^{J_{+}t^{\prime}}}{2\omega\sqrt{J_{+}^2+4\omega^2}}\cos 2\omega t^{\prime}\sin(2\omega t^{\prime}+\beta_{\omega})
\nonumber\\
&= C_1 \cos 4\omega t + C_2 \sin 4\omega t + C_3,
\label{Seq:y2solution}
\end{align}
where the coefficients $C_1$, $C_2$, and $C_3$ are given by
\begin{align}
C_1 &= \frac{J_{-}(-J_{+}^4-4\omega^2 J_{+}^2 - 8\omega^2 J_{-}^2 +J_{+}^2 J_{-}^2)}{4\omega(J_{+}^2+16\omega^2)(J_{+}^2+4\omega^2)}, \nonumber\\
C_2 &= \frac{J_{+}J_{-}(-2 J_{+}^2-8\omega^2+3 J_{-}^2)}{2(J_{+}^2+16\omega^2)(J_{+}^2+4\omega^2)}, \nonumber\\
C_3 &=\frac{J_{-}(-J_{+}^2-4\omega^2+J_{-}^2)}{4\omega(J_{+}^2+4\omega^2)}.
\end{align}
We note that, in the final expression of Eq.~\eqref{Seq:y2solution}, terms multiplied by exponential factor such as $e^{-J_{+}t}$ are omitted, as they become negligible in the limit $t\rightarrow\infty$.

Substituting the expressions for $Y_{\omega}^{(1)}$ and $Y_{\omega}^{(2)}$ into Eq.~\eqref{Seq:Xomega_epsilon}, and rearranging the terms by the order of $\varepsilon$, we  finally
obtain the following expansion for the dynamics of $X_\omega$ up to $O(\varepsilon^2)$:
\begin{align}
\dot{X}_{\omega} &= 2\omega (1+A_{\omega}\varepsilon^2) + \varepsilon (B_{\omega}\sin X_{\omega} + \omega C_{\omega}\cos X_{\omega}) \nonumber\\
&~~~+\varepsilon^2 (D_{\omega}\sin 2X_{\omega} + \omega  E_{\omega}\cos 2X_{\omega}) + O(\varepsilon^3), \label{Seq:Xomega_final}
\end{align}
where the coefficients $A_{\omega}$, $B_{\omega}$, $C_{\omega}$, $D_{\omega}$,and $E_{\omega}$
are all even functions of $\omega$, explicitly given by
\begin{align}
A_{\omega} &= \frac{J_{-}^2}{2(J_{+}^2+4\omega^2)}, \label{Seq:A} \\
B_{\omega} &= \frac{J_{+}(-J_{+}^2-4\omega^2+J_{-}^2)}{J_{+}^2+4\omega^2}, \label{Seq:B} \\
C_{\omega} &=-  \frac{2  J_{-}^2}{J_{+}^2+4\omega^2}, \label{Seq:C} \\
D_{\omega} &= - \frac{J_{-}^2 J_{+}(-J_{+}^4+4\omega^2 J_{+}^2 + 32\omega^4 + J_{+}^2 J_{-}^2-20\omega^2 J_{-}^2)}{2(J_{+}^2 + 4\omega^2)^2(J_{+}^2 + 16\omega^2)}, \label{Seq:D} \\
E_{\omega} &= - \frac{ J_{-}^2 (-4J_{+}^2 J_{-}^2+3J_{+}^4 + 12\omega^2 J_{+}^2 + 8\omega^2 J_{-}^2)}{(J_{+}^2 + 4\omega^2)^2 (J_{+}^2 + 16\omega^2)}. \label{Seq:E}
\end{align}
Note that all coefficients remain finite in the $\omega\to 0$  limit and do not diverge in the $\omega\to\infty$ limit, ensuring
regular behavior in the entire range of $\omega$.

Now, we compute $P_{\omega}^{d}(X_{\omega})$ up to the order $\varepsilon^2$, using Eq.~\eqref{Seq:Xomega_final} through the formulation in Eq.~\eqref{Seq:P_sameJK}.
A key observation is that the presence of the $\cos X_\omega$ and $\sin 2X_\omega$ terms in Eq.~\eqref{Seq:Xomega_final} explicitly breaks the $\pi$-reflection symmetry
of $P_{\omega}^{d}(X_{\omega})$.
This symmetry breaking implies that
 the dynamic contribution to the order parameter, $S_+^d$, may become nonzero even at  $O(\varepsilon^2)$.

To evaluate this contribution quantitatively,
we first calculate the normalization integral $\int_{-\pi}^{\pi} {\dot{X}}_{\omega}^{-1} dX_{\omega}$, which
is related to the effective average frequency $\tilde\omega$ for given $\omega$ in the long-time limit via
\begin{align}\label{Seq:mod1}
\int_{-\pi}^\pi \frac{dX_\omega}{\dot X_\omega} = \frac{2\pi}{\tilde\omega }~.
\end{align}
This yields the expression for the probability distribution function
\begin{align}\label{Seq:mod2}
P_\omega^d (X_\omega)=\frac{\tilde\omega}{2\pi {\dot X_\omega}}.
\end{align}
To compute the integral over $\omega$ in the expression for $S_+^d$ in Eq.~\eqref{Seq:Spd},
it is necessary to consider two distinct regimes based on the scale of $\omega$;
(i) $\omega \sim O(1)$ and (ii) $\omega \sim O(\varepsilon^z)$ with $0<z\leq 1$.

First, consider the case where $\omega \sim {O}(1)$. For convenience, we decompose  $\dot{X}_\omega$ into two parts, $F_1$ and $F_2$, given by
\begin{align}
F_1 &=2\omega(1+A_{\omega}\varepsilon^2)+B_{\omega}\varepsilon\sin X_{\omega}+
\omega C_{\omega}\varepsilon\cos X_{\omega}, \label{Seq:F1} \\
F_2 &=D_{\omega}\varepsilon^2\sin 2X_{\omega}+ \omega E_{\omega}\varepsilon^2
\cos 2X_{\omega}, \label{Seq:F2}
\end{align}
where we note that $F_1\sim O(1)$ and $F_2 \sim O(\varepsilon^2)$.
With this decomposition, the normalization integral
can be expanded in powers of the small parameter $F_2/F_1$ as
\begin{align}
\int_{-\pi}^{\pi} \frac{dX_{\omega}}{{\dot{X}}_{\omega}} &=\int_{-\pi}^{\pi}
\frac{dX_{\omega}}{F_1(1+F_2/F_1)} \nonumber\\
&=\int_{-\pi}^{\pi}dX_{\omega}\frac{1}{F_1}\Bigg(1-\frac{F_2}{F_1}+\frac{F_2^2}{F_1^2}+\cdots\Bigg) \nonumber \\
&=\int_{-\pi}^{\pi}\frac{dX_{\omega}}{F_1} - \int_{-\pi}^{\pi} dX_{\omega}\frac{F_2}{F_1^2}
+ O(\varepsilon^4). \label{Seq:F1F2}
\end{align}
The leading-order term can be evaluated exactly and yields
\begin{align}
\int_{-\pi}^{\pi}\frac{dX_{\omega}}{F_1}
=\frac{2\pi \cdot {\mbox{sgn}}(\omega)}{\sqrt{4\omega^2-(B_{\omega}^2+\omega^2 C_{\omega}^2-8\omega^2 A_{\omega})\varepsilon^2}}+O(\varepsilon^4). \label{Seq:Q1}
\end{align}
The second integral in Eq.~\eqref{Seq:F1F2} can also be evaluated as
\begin{align}
&\int_{-\pi}^{\pi} dX_{\omega}\frac{F_2}{F_1^2} \nonumber\\
	&=\varepsilon^2\int_{-\pi}^{\pi}dX_{\omega}\frac{(D_{\omega}\sin 2X_{\omega}+\omega E_{\omega}\cos 2X_{\omega})}{(2\omega)^2} +O(\varepsilon^3)\nonumber\\
	&= O(\varepsilon^4)~, \label{Seq:Q2}
\end{align}
where the lower-order contributions ($O(\varepsilon^2)$ and $O(\varepsilon^3)$ terms) vanish simply due to the periodicity of trigonometric functions in the integrand.
Combining Eqs.~(\ref{Seq:Q1}) and (\ref{Seq:Q2}) with the definition of $\tilde\omega$ in Eq.~\eqref{Seq:mod1}, we find
\begin{align}
\tilde\omega
=\frac{\sqrt{4\omega^2-(B_{\omega}^2+ \omega^2 C_{\omega}^2-8\omega^2A_{\omega})\varepsilon^2}}{ {\mbox{sgn}}(\omega)} + O(\varepsilon^4)~. \label{Seq:periodO1}
\end{align}
Note that $\tilde\omega$ is an odd function of $\omega$.

Now, for $\omega \sim  O(\varepsilon^z)$ with $0< z \leq 1$, we  decompose $\dot X_\omega$ into two parts, $F_1^\prime$ and $F_2^\prime$, which are given by
\begin{align}
F_1^\prime &= 2\omega + B_{\omega} \varepsilon \sin X_{\omega}, \label{Seq:F1_omega_epsilon}\\
F_2^\prime &= \omega C_{\omega}\varepsilon \cos X_{\omega} +  D_{\omega} \varepsilon^2 \sin 2X_{\omega}.
\label{Seq:F2_omega_epsilon}
\end{align}
In this decomposition, $F_1^\prime \sim O(\varepsilon^z)$ and $F_2^\prime \sim O(\varepsilon^{1+z})$. We neglect higher-order terms
such as those involving $A_\omega$ and $E_\omega$, as they do not affect the expansion up to $O(\varepsilon^2)$.
Following the same expansion scheme used  in Eq.~\eqref{Seq:F1F2},
we obtain
\begin{align}
	\int_{-\pi}^{\pi} \frac{dX_{\omega}}{{\dot{X}}_{\omega}} &=\int_{-\pi}^{\pi}\frac{dX_{\omega}}{F_1^\prime} - \int_{-\pi}^{\pi} dX_{\omega}\frac{F_2^\prime}{{F_1^\prime}^2}
	+ O(\varepsilon^{2-z}). \label{Seq:F1pF2p}
\end{align}
Note that we do not calculate the $O(\varepsilon^{2-z})$ term explicitly,  as it does not contribute to $S_+^d$ up to  $O(\varepsilon^2)$, as will be shown later.
The first integral in Eq.~\eqref{Seq:F1pF2p} can be evaluated exactly as
\begin{align}
\int_{-\pi}^{\pi}\frac{dX_{\omega}}{F_1^\prime}&=\int_{-\pi}^{\pi} \frac{dX_{\omega}}{2\omega+ B_{\omega} \varepsilon \sin X_{\omega}} = \frac{2\pi\cdot{\mbox{sgn}}(\omega)}{\sqrt{4\omega^2-B_\omega^2 \varepsilon^2}}.
\label{Seq:F1_1}
\end{align}
The second integration can be evaluated as
\begin{align}
&\int_{-\pi}^{\pi} dX_{\omega}\frac{F_2^\prime}{{F_1^\prime}^2}
=\omega C_{\omega} \varepsilon \int_{-\pi}^{\pi} \frac{\cos X_{\omega} dX_{\omega}}{(2\omega+B_{\omega}\varepsilon\sin X_{\omega})^2}
\nonumber\\
&\qquad +\varepsilon^2 D_{\omega}\int_{-\pi}^{\pi}\frac{\sin 2X_{\omega}dX_{\omega}}{(2\omega+B_{\omega}\varepsilon\sin X_{\omega})^2} =0. \label{Seq:F2_1}
\end{align}
From Eqs.~(\ref{Seq:F1_1}) and (\ref{Seq:F2_1}),  we find
\begin{align}
\tilde\omega = \frac{\sqrt{4\omega^2-B_{\omega}^2\varepsilon^2}}{{\mbox{sgn}(\omega)}} +O(\varepsilon^{2+z}). \label{Seq:periodOepsilon}
\end{align}
It is worth noting that  Eq.~\eqref{Seq:periodO1} remains valid in this case as well, since the additional contributions from the  $C_\omega$ and $A_\omega$
terms only affect the results at $O(\varepsilon^{2+z})$, which is beyond the accuracy required for the current analysis.
Using this result, the distribution function $P_\omega^d(X_\omega)$ is directly obtained through Eq.~\eqref{Seq:mod2}.
The dynamic contribution to the global ordering,  $S_{+}^d$ in Eq.~\eqref{Seq:Spd}, is now rewritten as
\begin{align}
S_{+}^d &=\int_{|\omega|>a\varepsilon} d\omega g(\omega) \int_{-\pi}^{\pi} dX_{\omega} e^{iX_{\omega}} P_{\omega}^d(X_{\omega}) \nonumber \\
&=\int_{a\varepsilon}^{\infty} d\omega g(\omega)
\int_{-\pi}^{\pi} dX_\omega e^{iX} {\tilde{P}}_{\omega}^d(X_\omega)~.
\label{Seq:Spd_Pd}
\end{align}
Here, ${\tilde{P}}_{\omega}^d(X_\omega) = P_{\omega}^{d}(X_\omega) + P_{-\omega}^{d}(X_\omega)$ for $\omega>0$ is expressed as  follows:
\begin{align}
{\tilde{P}}_{\omega}^d(X_\omega) = \frac{ \tilde\omega  G_1}{\pi (G_1^2 - G_2^2)}, \label{Seq:Ptilde}
\end{align}
where $G_1$ and $G_2$ are given by
\begin{align}
G_1 &=2\omega(1+A_{\omega}\varepsilon^2)+\omega C_{\omega}\varepsilon\cos X_{\omega}+\omega E_{\omega}\varepsilon^2 \cos 2X_{\omega}, \nonumber\\
G_2 &=B_{\omega}\varepsilon\sin X_{\omega}+D_{\omega}\varepsilon^2\sin 2X_{\omega}.
\end{align}

To evaluate the integral in Eq.~\eqref{Seq:Spd_Pd}, we must divide the integration range of
$\omega$ into two regions, as the asymptotic expansion of the integrand depends sensitively on the scale of $\omega$.
Accordingly, we separate the integration as follows:
\begin{align}
	S_{+}^d = H_1 + H_2,
\end{align}
where the terms $H_1$ and $H_2$ are given by
\begin{align}
	H_1 &=\int_{a\varepsilon}^{\delta \varepsilon^z} d\omega g(\omega) \frac{  \tilde\omega }{\pi}  \int_{-\pi}^{\pi} dX_{\omega}  \frac{e^{iX_{\omega}} G_1}{G_1^2-G_2^2}, \label{Seq:H1} \\
	H_2 &=\int_{\delta \varepsilon^z}^{\infty} d\omega g(\omega) \frac{ \tilde\omega }{\pi} \int_{-\pi}^{\pi} dX_{\omega}  \frac{e^{iX_{\omega}} G_1}{G_1^2 - G_2^2},
	\label{Seq:H2}
\end{align}
where $\delta$ denotes an  $O(1)$ constant.
It turns out that both integrands of $H_1$ and $H_2$ allows consistent expansions that are analytically integrable for $1/3<z<1/2$.
For explicit evaluation, we choose $z = 2/5$ as a representative value in the following.

First, we calculate $H_2$. After changing the integration variable as $\omega = \varepsilon^{\frac{2}{5}}y$, the integrand can be expanded up to  $O(\varepsilon^2)$ as follows:
\begin{align}
	&H_2 \approx \frac{\varepsilon^{\frac{2}{5}}}{\pi}\int_{\delta }^{\infty} dy g(\varepsilon^{\frac{2}{5}}y) \sqrt{ 4y^2-B_{\varepsilon^{2/5}y}^2\varepsilon^{\frac{6}{5}}}  \nonumber \\
	& \times \int_{-\pi}^{\pi} dX_{\omega}  e^{iX_{\omega}} \left[ \frac{1}{2y} - \frac{C_{\varepsilon^{2/5}y}\varepsilon}{4y} \cos X_{\omega} +\frac{B_{\varepsilon^{2/5}y}^2 \varepsilon^{\frac{6}{5}}}{8y^3}
\sin^2 X_{\omega} \right] \nonumber \\
	& = J_-^2 \varepsilon^{\frac{7}{5}}\int_{\delta }^{\infty} dy g(\varepsilon^{\frac{2}{5}}y)  \frac{\sqrt{ 4y^2-B_{\varepsilon^{2/5}y}^2\varepsilon^{\frac{6}{5}}}}{2y(J_+^2 + 4 \varepsilon^{\frac{4}{5}} y^2)},
	\label{Seq:H2_1}
\end{align}
where $B_{\varepsilon^{2/5}y}$ and $C_{\varepsilon^{2/5}y}$ are the values of
$B_{\omega}$ and $C_{\omega}$ in Eqs.~\eqref{Seq:B} and \eqref{Seq:C} for $\omega=\varepsilon^{2/5}y$.
As we retain terms up to $O(\varepsilon^2)$, the square root can be approximated as $\sqrt{ 4y^2-B_{\varepsilon^{2/5}y}^2\varepsilon^{\frac{6}{5}}} \approx 2y$. This allows us to further evaluate $H_2$ as
\begin{align}
	&H_2  = \frac{\gamma J_-^2}{\pi} \varepsilon^{\frac{7}{5}}\int_{\delta }^{\infty}  \frac{dy}{(\gamma^2 + \varepsilon^{\frac{4}{5}} y^2)(J_+^2 + 4 \varepsilon^{\frac{4}{5}} y^2)} \nonumber \\
	& = \frac{\gamma J_-^2}{\pi} \varepsilon
	\left[  \frac{ \pi (J_+ - 2\gamma) + 4 \gamma \tan^{-1}(\frac{2\delta \varepsilon^{\frac{2}{5}}}{J_+}) -2J_+ \tan^{-1}(\frac{\delta \varepsilon^{\frac{2}{5}}}{\gamma})}{2\gamma J_+(J_+^2 - 4\gamma^2)}
	\right] \nonumber \\
	&\approx \frac{J_{-}^2}{2J_{+}(J_{+}+2\gamma)}\varepsilon -\frac{J_{-}^2}{\gamma\pi J_{+}^2}\delta \varepsilon^{7/5}~. 
	\label{Seq:H2_final}
\end{align}

Next, we calculate $H_1$. Changing the integration variable as $\omega = \varepsilon y$ and expanding the integrand up to order $O(\varepsilon^2)$ yields
\begin{align}
	&H_1 \approx  \frac{2\varepsilon}{\pi}\int_{a}^{\delta \varepsilon^{-\frac{3}{5}}} dy g(\varepsilon y) \sqrt{ y^2-B_{\varepsilon y}^2 /4} \int_{-\pi}^{\pi} dX_{\omega} e^{iX_{\omega}} \nonumber\\
&\times \left[ \frac{2y}{4y^2-B_{\varepsilon y}^2 \sin^2 X_{\omega}} - \frac{y G_3 \varepsilon}{(4y^2-B_{\varepsilon y}^2 \sin^2 X_{\omega})^2}	
	 \right],
	\label{Seq:H1_1}
\end{align}
where $G_3 \equiv   -B_{\varepsilon y}(8D_{\varepsilon y}-B_{\varepsilon y}C_{\varepsilon y}) \cos X_\omega \sin^2 X_\omega  +4C_{\varepsilon y} y^2 \cos X_\omega   $.
Here, $B_{\varepsilon y}$, $C_{\varepsilon y}$, and $D_{\varepsilon y}$ represents the
values of $B_{\omega}$, $C_{\omega}$, and $D_{\omega}$ in Eqs.~\eqref{Seq:B}, \eqref{Seq:C}, and
\eqref{Seq:D} for $\omega=\varepsilon y$, respectively.
The first integral in Eq.~\eqref{Seq:H1_1} vanishes. Thus, $H_1$ can be rewritten as
\begin{align}
	&H_1 \approx  \frac{2\varepsilon^2}{\pi}\int_{a}^{\delta \varepsilon^{-\frac{3}{5}}} dy g(\varepsilon y) \sqrt{ y^2-B_{\varepsilon y}^2 /4}  \nonumber \\
	& \times \int_{-\pi}^{\pi} dX_{\omega}  e^{iX_{\omega}}
	\left[  \frac{ -y G_3 }{(4y^2-B_{\varepsilon y}^2 \sin^2 X_{\omega})^2}	
	\right].
	\label{Seq:H1_2}
\end{align}
As $\varepsilon^2$ is already included in the prefactor of $H_1$, all other terms with positive powers of $\varepsilon$ are of higher order than $O(\varepsilon^2)$, thus, can be neglected.
This allows us to approximate $B_{\varepsilon y} \approx - 2a$, $C_{\varepsilon y} \approx -2{J_{-}^2}/{J_{+}^2}$, $D_{\varepsilon y} \approx a{J_{-}^2}/{J_{+}^2}$, and
$g(\varepsilon y) \approx 1/(\pi \gamma)$, which leads to
\begin{align}
	H_1 &\approx  \frac{ \varepsilon^2 {J_{-}^2}}{\gamma \pi^2 {J_{+}^2} }
	\int_{a}^{\delta \varepsilon^{-\frac{3}{5}}} dy ~ y \sqrt{ y^2-a^2 } \int_{-\pi}^{\pi}
	\frac{ \cos X_{\omega} e^{iX_{\omega}} dX_{\omega} }{y^2-a^2 \sin^2 X}  \nonumber \\
	& =   \frac{ 2 \varepsilon^2  {J_{-}^2}}{3  a^2 \gamma \pi {J_{+}^2}} \left[ (d^2-a^2)^{\frac{3}{2} } - (d^3 -3a^2 d +2a^3)  \right],
	\label{Seq:H1_3}
\end{align}
where $d \equiv \delta \varepsilon^{-\frac{3}{5}}$. Therefore, by expanding Eq.~\eqref{Seq:H1_3} up to the order $\varepsilon^2$, we finally obtain
\begin{align}
	H_1 &\approx \frac{{J_{-}^2}}{\gamma\pi {J_{+}^2}}\delta \varepsilon^{7/5}
	- \frac{4a{J_{-}^2}}{3\pi\gamma{J_{+}^2}}\varepsilon^2. \label{Seq:H1_final}
\end{align}

Consequently, by replacing the expansion parameter $\varepsilon$ with $S_{+}$, the order parameter $S_{+}^d$ can be obtained by combining $H_1$ and $H_2$ from Eqs.~\eqref{Seq:H1_final} and \eqref{Seq:H2_final}, respectively, resulting in
\begin{align}
S_{+}^d = c_1 S_{+} - c_2 S_{+}^2. \label{Seq:Spd_c1c2}
\end{align}
where the coefficients $c_1$ and $c_2$ are given by
\begin{align}
c_1 = \frac{J_{-}^2}{2J_{+}(J_{+}+2\gamma)}
\quad \mbox{and} \quad c_2 =\frac{4a{J_{-}^2}}{3\pi\gamma{J_{+}^2}}. \label{Seq:c1c2}
\end{align}
Notably, the boundary terms that contain $\delta \varepsilon^{7/5}$ in $H_1$ and $H_2$ are exactly cancelled each other when $H_1$ and $H_2$ are combined.

By combining $S_{+}^s$ in Eq.~(\ref{Seq:Sps_JneqK_approx}) with
$S_{+}^d$ in Eq.~(\ref{Seq:Spd_c1c2}), the total order parameter $S_{+}$ is given by
\begin{align}
S_{+} &= S_{+}^s + S_{+}^d, \nonumber\\
 &= \bigg(\frac{a}{2\gamma}+c_1 \bigg) S_{+} - c_2 S_{+}^2 + {\cal{O}}(S_{+}^3).
\label{Seq:Sp_c1c2}
\end{align}
From Eq.~\eqref{Seq:Sp_c1c2}, keeping terms up to ${\cal{O}}(S_{+}^2)$, we obtain
\begin{align}
S_{+} = \frac{1}{c_2}\left(\frac{a}{2\gamma}+c_1-1 \right).
\end{align}
In terms of the parameters $J_\pm$, the order parameter $S_{+}$ can be expressed as
\begin{align}
S_{+} = \frac{3\pi}{8} \bigg(\frac{J_{+}^3}{J_{-}^2}\bigg) \frac{J_{+}^2-J_{-}^2-2\gamma(J_{+}+4\gamma)}{(J_{+}^2-J_{-}^2)(J_{+}+2\gamma)},
\end{align}
which can also be written in the scaling form near the transition point  as
\begin{align}
S_{+}\simeq \alpha_{+} (J_{+}-J_{+}^c)^{\beta_{+}}, \label{Seq:Sp_overall}
\end{align}
where $\alpha_{+}=\frac{\partial S_{+}}{\partial J_{+}}|_{J_+=J_+^c}$ (see the final subsection),
and the critical exponent is $\beta_{+}=1$, differing from the conventional Kuramoto model, which
has $\beta=1/2$.
The transition point $J_{+}^c$ is determined by the condition
\begin{align}
\frac{a}{2\gamma}+c_1-1=0, \label{Seq:threshold}
\end{align}
which yields
\begin{align}
J_{+}^c=\gamma+\sqrt{J_{-}^2 + (3\gamma)^2}. \label{Seq:Jpc}
\end{align}

Near the transition, the ratio of dynamic and static contributions is approximately given by
$S_+^d/S_+^s \approx (J_+-4\gamma)/(J_+ +4\gamma)$, indicating that while the influence of
dynamic oscillators increases with $J_+$ (equivalently with $J_-$), it remains bounded.

\subsection{Order parameter $S_{-}$ for $J_{-}\neq 0$}

Following a similar procedure, the order parameter $S_{-}$ can also be derived. In the limit $N \rightarrow \infty$, the order parameter $S_{-}^{s}$ for the static oscillators is given by
\begin{align}
S_{-}^s &= \int_{-aS_{+}}^{aS_{+}} e^{-i\sin^{-1} (\omega/(bS_-))} g(\omega)d\omega, \nonumber\\
&= \int_{-aS_{+}}^{aS_{+}}d\omega g(\omega)\sqrt{1-\bigg(\frac{\omega}{bS_{-}}\bigg)^2}.
\label{Seq:Zms}
\end{align}
Since $g(\omega)$ is an even function of $\omega$, the imaginary part disappears.
Therefore, the integration is reduced to
\begin{align}
S_{-}^s &= 2bS_{-} \frac{\gamma}{\pi}\int_{0}^{\frac{aS_{+}}{bS_{-}}} dx
\frac{\sqrt{1-x^2}}{(bS_{-}x)^2+\gamma^2} \nonumber\\
&=\frac{2a}{\pi\gamma}S_{+}+{\cal{O}}(S_+^3).
\label{Seq:Sms_JneqK}
\end{align}
The next step is to calculate the dynamic contribution $S_{-}^d$. Using Eq.~\eqref{Seq:Y_epsilon},
$Y-\Phi_-= Y_{\omega}=Y_{\omega}^{(1)}\varepsilon+Y_{\omega}^{(2)}\varepsilon^2$, the {\it time-dependent} order parameter $S_{-}^d (t)$ is given by
\begin{align}
S_{-}^d (t) &= \int_{|\omega|>aS_{+}} d\omega g(\omega)  e^{i(Y-\Phi_{-})}  \nonumber \\
&=\int_{aS_{+}}^{\infty} d\omega g(\omega) (e^{iY_{\omega}} + e^{iY_{-\omega}}). \label{Seq:Smd_time}
\end{align}
Note that the time-averaged value of $S_-^d(t)$ yields the order parameter $S_-^d$, i.e., $S_-^d = \langle S_-^d \rangle_t$, where $\langle \cdots \rangle_t$ denotes the time average.
The integrand of Eq.~\eqref{Seq:Smd_time} can be expanded as
\begin{align}
e^{iY_{\omega}}+e^{iY_{-\omega}} &= 2 - \frac{1}{2}\left({Y_{\omega}^{(1)}}^2 + {Y_{-\omega}^{(1)}}^2\right)\varepsilon^2  \nonumber\\
&+i(Y_{\omega}^{(1)}+Y_{-\omega}^{(1)})\varepsilon + i(Y_{\omega}^{(2)}+Y_{-\omega}^{(2)})\varepsilon^2 + {\cal{O}}(\varepsilon^3).
\end{align}
Averaging it over time with the explicit solution for $Y_\omega^{(1)}$ in Eq.~\eqref{Seq:y1solution}, we obtain
\begin{align}
\langle e^{iY_{\omega}} + e^{iY_{-\omega}} \rangle_t = 2-\frac{J_{-}^2 \varepsilon^2}{2(J_{+}^2+4\omega^2)} + O(\varepsilon^3), \label{Seq:timeAvg_eiY}
\end{align}
where $\langle \cos^2 2 \omega t \rangle_t = \frac{1}{2}$,
$\langle \cos 2 \omega t \rangle_t = 0$, $\langle \sin 2 \omega t \rangle_t = 0$, $\langle \cos 4 \omega t \rangle_t = 0$, and $\langle \sin 4 \omega t \rangle_t = 0$ are used.
Substituting Eq.~\eqref{Seq:timeAvg_eiY} into Eq.~\eqref{Seq:Smd_time}, we obtain
\begin{align}
S_{-}^d = \langle S_-^d \rangle_t  =2\int_{aS_{+}}^{\infty} d\omega g(\omega) -
\frac{1}{2}\int_{aS_{+}}^{\infty} d\omega g(\omega)\frac{J_{-}^2 \varepsilon^2}{J_+^2 + 4\omega^2} .  \label{Seq:Smd_time_Avg}
\end{align}
Using $g(\omega)=\frac{\gamma}{\pi}\frac{1}{\omega^2+\gamma^2}$, the first integral in Eq.~\eqref{Seq:Smd_time_Avg} is evaluated as
\begin{align}
\frac{2\gamma}{\pi} \int_{aS_{+}}^{\infty}\frac{d\omega}{\omega^2+\gamma^2}
=1 - \frac{2a}{\pi \gamma}S_{+} +  {\cal{O}}(S_{+}^3). \label{Seq:Smd_1st}
\end{align}
The result of the second integral in Eq.~\eqref{Seq:Smd_time_Avg}, evaluated up to $O(\varepsilon^2)$, leads to
\begin{align}
&\frac{1}{2}\int_{aS_{+}}^{\infty} \frac{J_{-}^2 \varepsilon^2}{J_+^2 + 4\omega^2} d\omega g(\omega)
\nonumber\\
&=\frac{\gamma J_-^2 \varepsilon^2}{2\pi} \int_{aS_{+}}^{\infty} \frac{d\omega}{(\omega^2+\gamma^2)(J_{+}^2+4\omega^2)} \nonumber\\
&= \frac{J_-^2 S_{+}^2}{4 J_{+}(J_{+}+2\gamma)} + {\cal{O}}(S_{+}^3),
\label{Seq:Smd_2nd}
\end{align}
where $\varepsilon$ is replaced with $S_{+}$.

By combining Eq.~(\ref{Seq:Smd_1st}) with (\ref{Seq:Smd_2nd}), the $S_{-}^d$ is given by
\begin{align}
S_{-}^{d} = 1-\frac{2a}{\pi\gamma}S_{+} - \frac{c_1}{2}S_{+}^2 + O(S_{+}^3).
\label{Seq:Smd_JneqK}
\end{align}
Then, using Eq.~(\ref{Seq:Sms_JneqK}) and Eq.~(\ref{Seq:Smd_JneqK}), the total order parameter
$S_{-} = S_-^s + S_+^d$ becomes
\begin{align}
S_{-}=1-\frac{c_1}{2}S_{+}^2 +O(S_+^3).
\label{Seq:Sm_withSp}
\end{align}
The absence of a linear term in $S_{+}$ is consistent with
the setting $S_{-}=1-\sigma\varepsilon^2$ (see Eq.~\eqref{Seq:Sm_epsilon}).
It is also worth noting that the dominant $O(1)$ contribution comes from the new dynamic mode associated with
bounded oscillations.

\begin{figure}
	\centering
        \includegraphics[width=0.95\columnwidth]{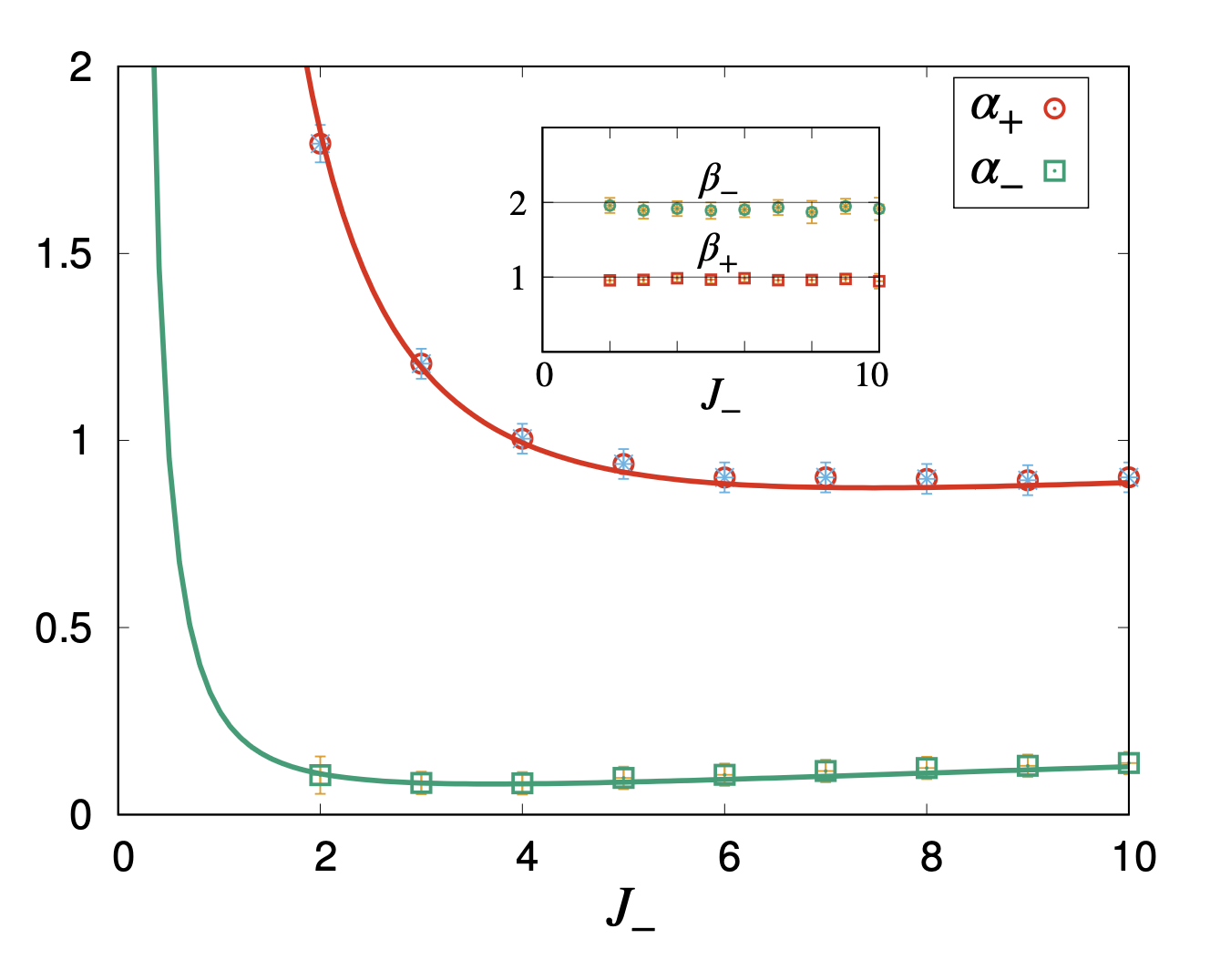}
	\caption{(Color online)
The amplitudes $\alpha_{\pm}$ and critical exponents $\beta_{\pm}$ are plotted as functions of $J_{-}$. Open red circles and green squares represent numerical data obtained from a system of size $N=10^5$, using a Lorentzian distribution with width $\gamma=1$.
Statistical errors are also indicated. The inset shows $\beta_{\pm}$ as a function of $J_{-}$. The solid red and green lines in the main panel correspond to Eqs.~(\ref{Seq:alphaplus}) and (\ref{Seq:alphaminus}), respectively, evaluated at $\gamma=1$.
\label{SMfig:alpha_beta}
}
\end{figure}

\subsection{Critical behavior of order parameters}
Using the scaling form for $S_{+}$ from Eq.~\eqref{Seq:Sp_overall} and substituting it into
Eq.~\eqref{Seq:Sm_withSp}, the behavior of the order parameter $S_{-}$ near the transition
point $J_{+}^c$ can be expressed as
\begin{align}
1-S_{-} =\alpha_{-}(J_{+}-J_{+}^c)^{\beta_{-}}, \label{Seq:Sm_alphaminus_betaminus}
\end{align}
where the critical exponent is $\beta_{-}=2\beta_{+}=2$, and
the amplitude $\alpha_{-}$ is given by
\begin{align}
\alpha_{-}=\frac{c_1}{2}\alpha_{+}^2.
\end{align}
The critical point $J_{+}^c$ is defined in Eq.~\eqref{Seq:Jpc}.
The amplitudes $\alpha_+$ and $\alpha_-$ are obtained by evaluating the derivatives at
$J= J_{+}^c$:
\begin{align}
\alpha_{+} &=\frac{\partial S_{+}}{\partial J_{+}}\Bigg|_{J_{+}=J_{+}^c} \nonumber\\
&=\frac{3\pi {J_{+}^c}^3(J_{+}^c-\gamma)}{8\gamma(J_{+}^c+2\gamma)^2({J_{+}^c}^2-16\gamma^2)}, \label{Seq:alphaplus} \\
\alpha_{-} &= \frac{\partial (1-S_{-})}{\partial J_{+}}\Bigg|_{J_{+}=J_{+}^c} \nonumber\\
&=\frac{9\pi^2 {J_{+}^c}^5 (J_{+}^c-\gamma)^2}{256\gamma^2 (J_{+}^c+2\gamma)^4 (J_{+}^c+4\gamma)({J_{+}^c}^2-16\gamma^2)}.
\label{Seq:alphaminus}
\end{align}
The critical exponents are summarized as $\beta_{+}=1$ and $\beta_{-}=2$.
These analytic results for $\alpha_\pm$ and $\beta_\pm$ are confirmed numerically in
Fig.~\ref{SMfig:alpha_beta}.
To evaluate the exponents $\beta_{\pm}$ numerically, we compute the local log-log slopes
between two consecutive data points, $J_{+}=J_{+}^{(1)}$ and $J_{+}=J_{+}^{(2)}$, using the following
expression:
\begin{align}
\tilde\beta_{\pm} = \frac{\ln S_{\pm}(J_{+}^{(1)}) -\ln S_{\pm}(J_{+}^{(2)})}{\ln (J_{+}^{(1)}-J_{+}^c) - \ln (J_{+}^{(2)}-J_{+}^c)}. \label{Seq:beta_pm}
\end{align}
Here, $S_{\pm}(J_{+}^{(1)})$ and $S_{\pm}(J_{+}^{(2)})$ denote the numerically
obtained values of $S_{\pm}$ at the neighboring values $J_{+}^{(1)}$ and $J_{+}^{(2)}$, respectively.
As $J_{+}^{(1)}$ and $J_{+}^{(2)}$ approach $J_{+}^c$, the estimated exponent $\tilde\beta_{\pm}$ converges
to the true value $\beta_\pm$ in the large-$N$ limit.
Once $\beta_{\pm}$ are determined, the corresponding amplitudes $\alpha_{\pm}$
can be calculated using the following relations:
\begin{align}
\alpha_{+} &= S_{+}(J_{+}-J_{+}^c)^{-\beta_{+}}, \label{Seq:alphaplus_from_Sp} \\
\alpha_{-} &= (1-S_{-})(J_{+}-J_{+}^c)^{-{\beta_{-}}}. \label{Seq:alphaminus_from_Sm}
\end{align}
We note that the expressions for $\alpha_{\pm}$ yield the same results as those presented in
Eqs.~\eqref{Seq:alphaplus} and ~\eqref{Seq:alphaminus} (see Fig.~\ref{SMfig:alpha_beta}).

\subsection{Stability analysis of static solution for $J_{-}\neq 0$}
In this subsection we briefly show the linear stability of the
static solution, $(X^s, Y^s)$, as given by Eqs.~(\ref{Seq:Xs}) and (\ref{Seq:Ys}), valid for $|\omega|\leq aS_+$.
To assess the stability of $(X^s, Y^s)$, we introduce small perturbations $\delta X, \delta Y \ll 1$ and analyze how these perturbations evolve. Substituting $X = X^s + \delta X$ and $Y = Y^s + \delta Y$ into Eqs.~(\ref{Seq:X}) and (\ref{Seq:Y}) and expanding the equations up to the orders $O(\delta X)$ and $ O(\delta Y)$, we obtain
\begin{equation}
\left(
\begin{matrix}
\delta \dot{X} \\
\delta \dot{Y}
\end{matrix}
\right)
=\textsf M
\left(
\begin{matrix}
\delta X \\
\delta Y
\end{matrix}
\right)
= -
\left(
\begin{matrix}
J_{+}S_{+} \tilde a & J_{-}S_{-} \tilde b\\
J_{-}S_{+} \tilde a & J_{+}S_{-} \tilde b
\end{matrix}
\right)
\left(
\begin{matrix}
\delta X \\
\delta Y
\end{matrix}
\right),
\end{equation}
where $\tilde a \equiv \sqrt{1-\frac{\omega^2}{a^2 S_{+}^2}}$ and $\tilde b \equiv \sqrt{1-\frac{\omega^2}{b^2 S_{-}^2}}$.
If we denote $\lambda_1$ and $\lambda_2$ as two eigenvalues of the matrix $\textsf M$, the stable condition of the static solution is given by ${\rm Re}(\lambda_1), {\rm Re}(\lambda_2) <0$. This condition is achieved when the following inequalities are satisfied:
\begin{align}
{\rm{Tr}} \textsf M &= \lambda_1 + \lambda_2 =  -J_{+}S_{+} \tilde a - J_{+}S_{-} \tilde b  <0, \label{Seq:TrM} \\
{\rm{det}} \textsf M &= \lambda_1 \lambda_2 = (J_+^2 - J_-^2)S_+ S_- \tilde a \tilde b >0. \label{Seq:detM}
\end{align}
Inequalities~\eqref{Seq:TrM} and \eqref{Seq:detM} imply that $J_{+}>0$ and $J_{+}>J_{-}$.
Returning to the validity of the static solutions in Eqs.~\eqref{Seq:Xs} and \eqref{Seq:Ys},
these conditions ensure that $a,b>0$.

	
	
%
	




\section*{References}

\begin{thebibliography}{99}
\bibitem{Winfree}
A. T. Winfree, {\it{The Geometry of Biological Time}} (Springer, 2001).

\bibitem{Crawford}
J. D. Crawford, J. Stat. Phys. {\bf 74}, 1047 (1994); Phys. Rev. Lett. {\bf 74}, 4341 (1995);
J. D. Crawford and K. T. R. Davies, Physica D {\bf 125}, 1 (1999).

\bibitem{Kuramoto}
Y. Kuramoto, in {\it{International Symposium on Mathematical Problems in Theoretical Physics}},
edited by H. Araki, Springer Lecture Notes Phys., Vol. {\bf 39} (Springer, New York, 1975), p.420;
{\it{Chemical Oscillations, Waves, and Turbulence}} (Springer, Berlin, 1984);
Y. Kuramoto and I. Nishikawa, J. Stat. Phys. {\bf 49}, 569 (1987).

\bibitem{Strogatz}
S. H. Strogatz and R. Mirollo, J. Stat. Phys. {\bf 63}, 613 (1991);
S. H. Strogatz, R. Mirollo, and P. C. Matthews, Phys. Rev. Lett. {\bf 68}, 2730 (1992);
M. K. S. Yeung and S. H. Strogatz, Phys. Rev. Lett. {\bf 82}, 648 (1999);
S. H. Strogatz, Physica D {\bf 143}, 1 (2000).

\bibitem{StrogatzBook}
S. H. Strogatz, {\it{Sync: The Emerging Science of Spontaneous Order}} (Hyperion, New York, 2003).

\bibitem{Pikovsky}
A. Pikovsky, M. Rosenblum, and J. Kurths, {\it{Synchronization: A Universal Concept in Nonlinear Sciences}} (Cambridge University Press, Cambridge, UK, 2003).

\bibitem{Daido}
H. Daido, Prog. Theor. Phys. {\bf 88}, 1213 (1992);
H. Daido, Phys. Rev. Lett. {\bf 73}, 760 (1994).

\bibitem{Acebron}
J. A.  Acebron, L. L. Bonilla, C. J. P. Vicente, F. Ritott, and R. Spigler, Rev. Mod. Phys. {\bf{77}}, 137 (2005).

\bibitem{Buck}
J. Buck and E. Buck, Science {\bf 159}, 1319 (1968).

\bibitem{Peskin}
C. S. Peskin, Mathematical Aspects of Heart Physiology (Courant Institute of Mathematical Sciences, New York, 1975), p.268.

\bibitem{Gray}
C. M. Gray, P. Konig, A.K. Engel, and W. Singer, Nature (London) {\bf 338}, 334 (1989).

\bibitem{StrogatzCircadianClock}
C. Liu, D. R. Weaver, S. H. Strogatz, and S. M. Reppert, Cell {\bf 91}, 855 (1997);
S. H. Strogatz, R. E. Kronauer, and C. A. Czeisler, Am. J. Physiol. Regul. Integr. Comp. Physiol. {\bf{253}}, R172-R178 (1987).

\bibitem{Hopfield}
J. J. Hopfield and A. V. M. Herz, Proc. Natl. Acad. Sci.  U.S.A. {\bf{92}}, 6655 (1995).

\bibitem{Kirst}
C. Kirst, T. Geisel, and M. Timme, Phys. Rev. Lett. {\bf{102}}, 068101 (2009).

\bibitem{Ott}
E. Ott and T. M. Antonsen, Chaos {\bf 27}, 051101 (2017).

\bibitem{Arenas}
A. Arenas, A. D. -Guilera, J. Kurths, Y. Moreno, C. Zhou, Phys. Rep. {\bf 469}, 93 (2008).

\bibitem{Neda}
Z. N{\'e}da, E. Ravasz, T. Vicsek, Y. Brechet, and A. L. Barab{\'a}si, Phys. Rev. E {\bf 61}, 6987 (2000).







\bibitem{KM_Kij}
H. Daido, Phys. Rev. Lett. {\bf 68}, 1073 (1992);
J. C. Stiller and G. Radons, Phys. Rev. E {\bf 58}, 1789 (1998);
H. Daido, Phys. Rev. E {\bf 61}, 2145 (2000);
J. C. Stiller and G. Radons, Phys. Rev. E {\bf 61}, 2148 (2000);
B. O.-L{\"o}ffler and S. H. Strogatz, Phys. Rev. Lett. {\bf 120}, 264102 (2018).

\bibitem{KM_bimodal}
E. A. Martens, E. Barreto, S. H. Strogatz, E. Ott, P. So, and T. M. Antonsen, Phys. Rev. E {\bf 79}, 026204 (2009).

\bibitem{KM_Ki}
H. Hong and S. H. Strogatz, Phys. Rev. Lett. {\bf 106}, 054102 (2011); Phys. Rev. E {\bf 84}, 046202 (2011); Phys. Rev. E {\bf 85}, 056210 (2012).

\bibitem{KM_local}
H. Hong, H. Park, and M. Y. Choi, Phys. Rev. E {\bf 72}, 036217 (2005);
H. Hong, H. Chat\'e, H. Park and L.-H. Tang, Phys, Rev. Lett. {\bf 99}, 184101 (2007).

\bibitem{KM_thermal}
H. Sakaguchi, S. Shinomoto, and Y. Kuramoto, Prog. Theor. Phys. {\bf 79}, 600 (1988);
S. -W. Son and H. Hong, Phys. Rev. E {\bf 81}, 061125 (2010);
I. V. Tyulkina, D. S. Goldobin, L. S. Klimenko, and A. Pikovsky, Phys. Rev. Lett. {\bf 120}, 264101 (2018).

\bibitem{KHS17}
K. P. O’Keeffe, H. Hong, S. H. Strogatz, Nature comm. {\bf{8}}, 1 (2017).

\bibitem{Kevin22}
K. P. O'Keeffe, S. Ceron, and K. Petersen Phys. Rev. E {\bf 105}, 014211 (2022); {\color{black}{Z. G. Nicolaou, D. Eroglu, and A. E. Motter, 
Phys. Rev. X {\bf 9}, 011017 (2019).}}

\bibitem{Yoon22}
S. Yoon, K. P. O'Keeffe, J.F.F. Mendes, and A. V. Goltsev, Phys. Rev. Lett. {\bf{129}}, 208002 (2022).

\bibitem{Hong23}
H. Hong, K. P. O’Keeffe, J. S. Lee, and H. Park, Phys. Rev. Research {\bf{5}}, 023105 (2023).


{\color{black} \bibitem{corr} Introducing correlations between $\omega_i$ and $v_i$  generates distinct frequency distributions (or distribution widths) for the transformed variables, $X_i$ and $Y_i$. In the case of perfect correlation, the corresponding distribution width associated with the $Y$ variable vanishes, leading to zero intrinsic frequency for all oscillators (see Eq.~\eqref{eq:Y}).}

\bibitem{WS}
S. Watanabe and S. H. Strogatz, Phys. Rev. Lett. {\bf 70}, 2391 (1993); Physica D {\bf 74}, 197 (1994);
J. Um, H. Hong, and H. Park, Sci. Rep. {\bf 14}, 6816 (2024).




\bibitem{ref:SM}
See supplementary material for additional derivations and figures.

\bibitem{Heun}
See, e.g., R.L. Burden and J.D. Faires, \textit{Numerical Analysis}
(Brooks/Cole, Pacific Grove, 1997), p.280.

\bibitem{LC} D. Andrienko, J. Molecular Liquids {\bf 267}, 520 (2018).
\bibitem{AM} V. Venkatesh, N. de G. Sousa, and A. Doostmohammadi, J. Phys. A: Math. Theor. {\bf 58}, 263001 (2025).



%
%
%
%
%
%
%
%

%
%
%
%
%
%
%
%
%
%
\end{thebibliography}
\end{document}